\documentclass[manuscript]{aastex631}
\usepackage{amsmath}
\usepackage{amssymb}

\newcommand{\kl}[1]{\left({#1}\right)}
\newcommand{\klm}[1]{\left[{#1}\right]}
\newcommand{\partt}[1]{\frac{\partial {#1}}{\partial t}}

\newcommand{\boldE}{\boldsymbol{E}}
\newcommand{\boldB}{\boldsymbol{B}}
\newcommand{\boldu}{\boldsymbol{u}}
\newcommand{\boldv}{\boldsymbol{v}}

\newcommand{\boldb}{\boldsymbol{b}}
\newcommand{\boldj}{\boldsymbol{j}}

\newcommand{\boldkappa}{\boldsymbol{\kappa}}
\newcommand{\boldsigma}{\boldsymbol{\sigma}}

\begin{document}

\title{Magnetic Energy Conversion in MHD: Curvature Relaxation and 
Perpendicular Expansion of Magnetic Fields}

\author[0000-0003-1134-3909]{Senbei Du}
\affiliation{Los Alamos National Laboratory, Los Alamos, NM 87545, USA}

\author[0000-0003-3556-6568]{Hui Li}
\affiliation{Los Alamos National Laboratory, Los Alamos, NM 87545, USA}

\author[0000-0002-4305-6624]{Xiangrong Fu}
\affiliation{New Mexico Consortium, Los Alamos, NM 87544, USA}

\author[0000-0003-3886-0383]{Zhaoming Gan}
\affiliation{New Mexico Consortium, Los Alamos, NM 87544, USA}

\author[0000-0002-4142-3080]{Shengtai Li}
\affiliation{Los Alamos National Laboratory, Los Alamos, NM 87545, USA}

\begin{abstract}

The mechanisms and pathways of magnetic energy conversion are an 
important subject for many laboratory, space and astrophysical systems.
Here, we present a perspective on magnetic energy conversion in MHD 
through magnetic field curvature relaxation (CR) and perpendicular expansion (PE)
due to magnetic pressure gradients,
and quantify their relative importance in two representative cases,
namely 3D magnetic reconnection and 3D kink-driven instability in an astrophysical jet. 
We find that the CR and PE processes have different temporal
and spatial evolution in these systems. 
The relative importance of the two processes tends to reverse as the system enters
the nonlinear stage from the instability growth stage.
Overall, the two processes make comparable contributions to the magnetic energy conversion
with the PE process somewhat stronger than the CR process.
We further explore how these energy conversion terms can be related 
to particle energization in these systems.

(This is the Accepted Manuscript version of an article accepted for publication in \textit{The Astrophysical Journal}.
IOP Publishing Ltd is not responsible for any errors or omissions in this version of the manuscript or any version derived from it.)

\end{abstract}

\section{Introduction}

The energization and heating of plasma is a universal phenomenon 
in laboratory, space physics 
and astrophysics, such as in the solar corona, solar wind, disks around 
active galactic nuclei, astrophysical jets, etc.
Magnetic energy is frequently found to be the free energy source for energizing 
plasma in these systems and thus the conversion of magnetic energy is a
key element for understanding these plasma systems
\citep[e.g.,][]{Parker1979, Colgate2001, Kronberg2001}.
Observations of many magnetically dominated astrophysical environments and 
systems have demonstrated fast magnetic energy release and strong particle acceleration.
Blazar jets powered by supermassive black holes can exhibit $\sim$ minutes variations
in TeV emissions \citep{Aharonian2007}
and gamma-ray flares from Crab nebula have been observed as well
\citep{Abdo2011}.

For example, one commonly considered case is magnetic reconnection 
as one of the most important 
magnetic energy conversion processes. 
The rapid change in magnetic field connectivity converts magnetic energy 
stored in anti-parallel magnetic field into plasma energy
\citep[e.g.,][]{Sweet1958, Parker1957, Parker1963, Priest2007, Birn2012} and
can lead to efficient particle acceleration and heating
\citep[e.g.,][]{Drake2006, Li2019}.
In addition, it is well established that magnetic reconnection occurring in global-scale 
current sheets can contribute to the energy release in solar flares 
and the Earth magnetosphere \citep{Dungey1961}.
More recently, \citet{Ripperda2020} demonstrate with general-relativistic resistive
MHD simulations that magnetic reconnection can generate plasmoids and explain black hole flares.
Another important case is the kink-instability enabled magnetic energy conversion. 
In astrophysical jets, the free magnetic energy can also be 
stored in large-scale helical magnetic fields and is released 
due to kink instability \citep[e.g.,][]{Li2006, Nakamura2007,
Mizuno2009, Zhang2017}.
The third case that is often studied is the magnetized turbulence
in which the injected magnetic energy can be converted to plasma energy as well. 
Here, the compression effect has been discussed in the context of 
single-fluid MHD and it is especially important for the plasma 
heating \citep{Birn2012, Du2018}. In particular, it was proposed 
\citep[e.g.,][]{Yang2016, Matthaeus2020} that
the electromagnetic energy interconverts with the flow kinetic energy
via the $\boldj\cdot\boldE$ term and the ``pressure-strain" interaction
converts energy between flow kinetic energy and the internal energy. 
The pressure-strain interaction consists both a compressive part
$- p \nabla \cdot \boldu$
[pressure dilatation, \cite{Aluie2012}] and a ``Pi-D'' term
which is the product of a traceless pressure tensor and the velocity
shear tensor, though its robustness 
may continue to be under scrutiny \citep{Du2020}.

Rapid progress has also been made in the area of plasma kinetic studies of 
magnetic reconnection and the associated 
particle energization along with theoretical developments 
\citep[e.g.,][]{Sironi2014, Guo2014, Dahlin2014, Zank2014, Werner2016,
Li2015, Li2017, Li2018, Li2019, leRoux2015, leRoux2018, Lazarian2020}. In particular, 
\citet{Li2017} constructed the electric current from various particle 
drift motions and evaluate the particle energization due to drifts in kinetic 
particle-in-cell (PIC) simulations of magnetic reconnection. 
While curvature drift is found to be the dominant particle acceleration mechanism, 
other drifts also have some minor contributions. Of particular importance 
is the gradient (or grad-B) drift, which usually has an decelerating effect 
and counteracts with the curvature drift acceleration.
\citet{Li2018} further show that combining the various drift acceleration terms 
yields an expression for the energization that can be interpreted as the 
summation of fluid compression, shear, and inertial energization effects.
In addition to the first-principle kinetic studies, 
the usage of the test particle approach
in reconnection and kink configurations 
\citep[e.g.,][]{Kowal2012, Medina2021} has also 
yielded useful understanding in particle energization processes.

Connecting the microscopic particle motion and acceleration 
processes with the macroscopic magnetic energy conversion processes
remains a long-standing challenge. 
\citet{Beresnyak2016} discussed the conversion 
of magnetic energy in MHD and its relation to the
first-order Fermi particle acceleration. Their analysis 
suggests that the energy transfer from magnetic field 
to kinetic motions is inherently related to the curvature drift acceleration.
\citet{Beresnyak2016} also discuss the distinction between 
different types of turbulence.
Specifically, the energization in decaying magnetic 
turbulence and magnetic reconnection driven turbulence 
are compared against each other using incompressible 
MHD simulations.
While both systems see a conversion of magnetic 
energy into kinetic energy, the magnetic reconnection case 
is found to have a stronger energization. This is likely due 
to the larger free magnetic energy that is available in 
the reconnection case.
In terms of the curvature drift, one may argue that 
the magnetic field posesses 
``good'' curvature that helps the release of 
magnetic energy and particle acceleration during reconnection.
However, as we will show in this paper, the conclusions made by \citet{Beresnyak2016} are contingent on the incompressible 
nature of the flow and need to be revisited for a general 
compressible plasma. 

In this paper, we will present detailed analyses of 
the magnetic energy conversion processes in two typical 
configurations, reconnection and kink, using 3D MHD simulations. 
We show that two dominant processes can be identified in regulating
the magnetic energy conversion. We will present results from 
the theoretical analysis in \S \ref{sec:theory} 
and the numerical analysis in \S \ref{sec:simulation}.
Summary and conclusions are given in \S \ref{sec:summary}.

\section{Energy conversion in MHD}\label{sec:theory}

\subsection{Two processes in magnetic energy conversion}

The ideal MHD energy equations, including magnetic energy and plasma energy, are
\begin{equation}\label{eq:debdt}
  \partt{}\kl{\frac{B^2}{8\pi}} + \nabla\cdot\klm{\frac{-(\boldu\times\boldB)\times\boldB}{4\pi}} = -\boldj\cdot\boldE;
\end{equation}
\begin{equation}\label{eq:tot}
  \partt{}\kl{\frac{1}{2}\rho u^2 + \frac{p}{\gamma - 1}} + \nabla\cdot\klm{\kl{\frac{1}{2}\rho u^2 + \frac{\gamma p}{\gamma - 1}}\boldu} = \boldj\cdot\boldE.
\end{equation}
We use the CGS Gaussian unit system throughout the paper.
As usual, the quantities in the equations are defined as magnetic field $\boldB$, electric field $\boldE$, current density $\boldj$, plasma mass density $\rho$, flow velocity $\boldu$, thermal pressure $p$, and adiabatic incex $\gamma$.
The work done by the electric field $\boldj\cdot\boldE$ converts the magnetic energy to the plasma energy, and thus contributes to the energization of plasma. It can be expanded as
\begin{equation}\label{eq:jdote}
  \boldj\cdot\boldE = -(\boldu\times\boldB)\cdot\frac{\nabla\times\boldB}{4\pi} = \frac{1}{4\pi}\boldu\cdot(\boldB\cdot\nabla\boldB) - \boldu\cdot\nabla\kl{\frac{B^2}{8\pi}}.
\end{equation}
We further expand the first term on the right hand side of Equation \eqref{eq:jdote} as
\begin{equation}\label{eq:ubgb}
  \frac{1}{4\pi}\boldu\cdot(\boldB\cdot\nabla\boldB) = \frac{1}{4\pi}(\boldu\cdot\boldB)(\boldb\cdot\nabla B) + \frac{B^2}{4\pi}\boldu\cdot\boldkappa,
\end{equation}
where the magnetic field curvature is defined as $\boldkappa = \boldb\cdot\nabla\boldb$ and $\boldb = \boldB/B$ is the unit vector tangent to the magnetic field. 
Combining these terms, we can re-cast Equation (\ref{eq:jdote}) as
\begin{equation}\label{eq:jdote2}
  \boldj\cdot\boldE = \frac{B^2}{4\pi}\boldu\cdot\boldkappa - \boldu\cdot\nabla_{\perp}\kl{\frac{B^2}{8\pi}} 
  ~.
\end{equation}
Here, we denote the gradients parallel and perpendicular to 
the magnetic field as $\nabla_{\parallel} = \boldb(\boldb\cdot\nabla)$ 
and $\nabla_{\perp} = \nabla - \nabla_{\parallel}$.
The equation shows that the total magnetic energy transfer in Equation (\ref{eq:debdt}) 
can be  decomposed into two parts. 
Physically, the first term corresponds to the relaxation of high magnetic field 
line tension -- curvature relaxation (CR) when the flow velocity is along the 
curvature of magnetic field, which releases magnetic energy. 
The second term corresponds to the perpendicular expansion (PE) 
of the magnetic energy, which can drive flows in the opposite direction of
the magnetic energy gradient, releasing magnetic energy. 

A complementary view is expressed via plasma energization 
by several previous studies \citep[e.g.,][]{Birn2012, Du2018} where 
Equation \eqref{eq:tot} can rewritten as 
the plasma flow energy and thermal energy evolution,
\begin{equation}\label{eq:bulk}
  \partt{}\kl{\frac{1}{2}\rho u^2} + \nabla\cdot\kl{\frac{1}{2}\rho u^2 \boldu} = \boldj\cdot\boldE - \boldu\cdot\nabla p~~;
\end{equation}
\begin{equation}\label{eq:thermal}
  \partt{}\kl{\frac{p}{\gamma - 1}} + \nabla\cdot\kl{\frac{\gamma p}{\gamma - 1}\boldu} = \boldu\cdot\nabla p~~.
\end{equation}
These equations suggest that the work done by the electric field $\boldj\cdot\boldE$ 
contributes strictly to the bulk acceleration in ideal MHD and 
heating is only facilitated by compression $\boldu\cdot\nabla p$.

\subsection{The role of fluid compression and shear}

Another way to express $\boldj\cdot\boldE$ is by using the 
perpendicular velocity $\boldu_{\perp}$ in the place of the 
total velocity $\boldu$ since the parallel velocity does not 
contribute to the electric field. It can be shown that the 
curvature drift term is related to the flow shear as
\[ \frac{B^2}{4\pi}\boldu\cdot\boldkappa = -\frac{B^2}{4\pi} \boldb\boldb:\nabla\boldu_{\perp}, \]
and that the total $\boldj\cdot\boldE$ is 
decomposed into the sum of perpendicular expansion and shear,
\begin{equation}
  \boldj\cdot\boldE = -\boldu_\perp \cdot\nabla\kl{\frac{B^2}{8\pi}} - \frac{B^2}{4\pi} \boldb\boldb:\nabla\boldu_{\perp},
\end{equation}
or
\begin{equation}\label{eq:jdote3}
  -\boldj\cdot\boldE = \nabla\cdot\kl{\frac{B^2}{8\pi}\boldu_{\perp}} - \frac{B^2}{8\pi}\frac{\nabla\cdot\boldu_{\perp}}{3} + \frac{B^2}{4\pi}\boldb\boldb:\boldsigma,
\end{equation}
where $\sigma_{ij} = (1/2)(\partial_i u_{\perp j} + \partial_j u_{\perp i} - 
2 \nabla\cdot\boldu_{\perp} \delta_{ij} / 3)$ is the shear tensor ($\delta_{ij}$ is the Kroneker delta).
Equation \eqref{eq:jdote3} is very reminiscent of the results 
obtained by \citet{Li2018}, who show from the Vlasov equation that 
in the limit of a gyrotropic (aka CGL) pressure 
\citep{Chew1956, Hunana2019}, 
the plasma energization in the perpendicular direction $\boldj_{\perp} \cdot \boldE_{\perp}$
can be expressed as the sum of 
a compression term, a shear term, and an inertial term
[see their Equation (9) in \citet{Li2018} where
$\boldj_{\perp}$ is the perpendicular component of the
current density w.r.t to the local magnetic field and 
such a current is produced by various particle motions and drifts]. 
Similarly in our Equation \eqref{eq:jdote3}, the first term
on the right does not contribute to magnetic energy conversion,
the second and third terms correspond to magnetic energy conversion
via flow expansion/compression and flow shear, respectively. 

The fact that Equations \eqref{eq:jdote2} and \eqref{eq:jdote3} are
mathematically equivalent suggests that both CR and PE processes
are mixed together in the shear term in Equation \eqref{eq:jdote3}.  
In this paper, we opt to use Equation \eqref{eq:jdote2} as our
primary approach to analyze the magnetic conversion processes.

\section{MHD simulations}\label{sec:simulation}

In this section, we demonstrate how magnetic
energy conversion occurs in different types of systems, 
namely magnetic reconnection and kink-unstable jet.
We have performed a set of 3D ideal MHD simulations 
using the \textit{Athena++} code \citep{Stone2008, Stone2020} and the \textit{LA-COMPASS} code
to study these systems. We focus on the temporal and spatial evolution
of the main terms described in Equation \eqref{eq:jdote2}. 

\subsection{Magnetic reconnection}\label{sec:reconnection}

The magnetic reconnection simulation is initialized with force-free current sheets 
with the following magnetic field configuration,
\begin{equation}
  B_x = B_0 \tanh\left[\frac{d}{\pi L}\sin\left(\frac{\pi z}{d}\right)\right];\quad B_y = B_0\sqrt{1 + \left(\frac{B_g}{B_0}\right)^2 - \tanh^2\left[\frac{d}{\pi L}\sin\left(\frac{\pi z}{d}\right)\right]};\quad B_z = 0.
\label{eq:rec_init}
\end{equation}
Here, $B_0$ is the strength of the upstream reconnecting magnetic field; 
$B_g$ is the guide field strength; $L$ is the half thickness of the current sheets; 
and $d$ is the distance between two adjacent current sheets.
We use a box size of $2\pi$ in all three directions for our simulation, 
resolved by $512^3$ cells. The boundary conditions are periodic in all directions.
We choose the parameters $B_0 = 1$, $B_g = 0$, $L = 0.05$, and $d = \pi$ 
so that there are two current sheets initially. 
The initial density and temperature profiles are both uniform. 
The simulation is normalized such that the initial Alfv\'en speed $V_A = 1$ and the initial plasma
$\beta = 0.2$ (ratio between the thermal pressure and magnetic pressure). 
An adiabatic equation of state is used with an adiabatic index $=5/3$.
The characteristic Alfv\'en time will be $\tau_A = 2\pi$. 
The current sheets are perturbed initially by random Alfv\'enic fluctuations 
that propagate in the $xy$-plane. We introduce 100 random sine waves with 
transverse velocity and magnetic fluctuations $\delta V_z$ and $\delta B_z$ 
that are correlated in an Alfv\'enic fashion ($\delta V_z = \pm \delta B_z$).
The total perturbation rms amplitude is $\sim 8\%$ and the cross helicity 
is close to zero. The perturbation is restricted to long-wavelength fluctuations 
with the wavelength equal or longer than a third of the box size.

We note that although we use ideal MHD without explicit resistivity and viscosity in the simulation,
magnetic reconnection can still occur due to numerical diffusivity at the grid scale.
In a high-Lundquist number plasma typical for space and astrophysical environments, the energy conversion
is dominated by ideal MHD process and resistive heating can generally be neglected at large scales.
For magnetic reconnection, the nonideal physics is important in the diffusion region where the magnetic field lines break,
but the energy conversion process for the entire current sheet system is not sensitive to the nonideal physics.
We do caution that due to the limited spatial resolution, our simulation corresponds to a moderate Lundquist number
on the order of hundreds. It is unclear at this point how our results hold in higher Lundquist number regimes
where the plasmoid instability becomes dominant \citep{Loureiro2007, Bhattacharjee2009, HuangY2010, YangL2020}.

\begin{figure}[!htp]
  \centering
  \includegraphics[width=0.5\linewidth]{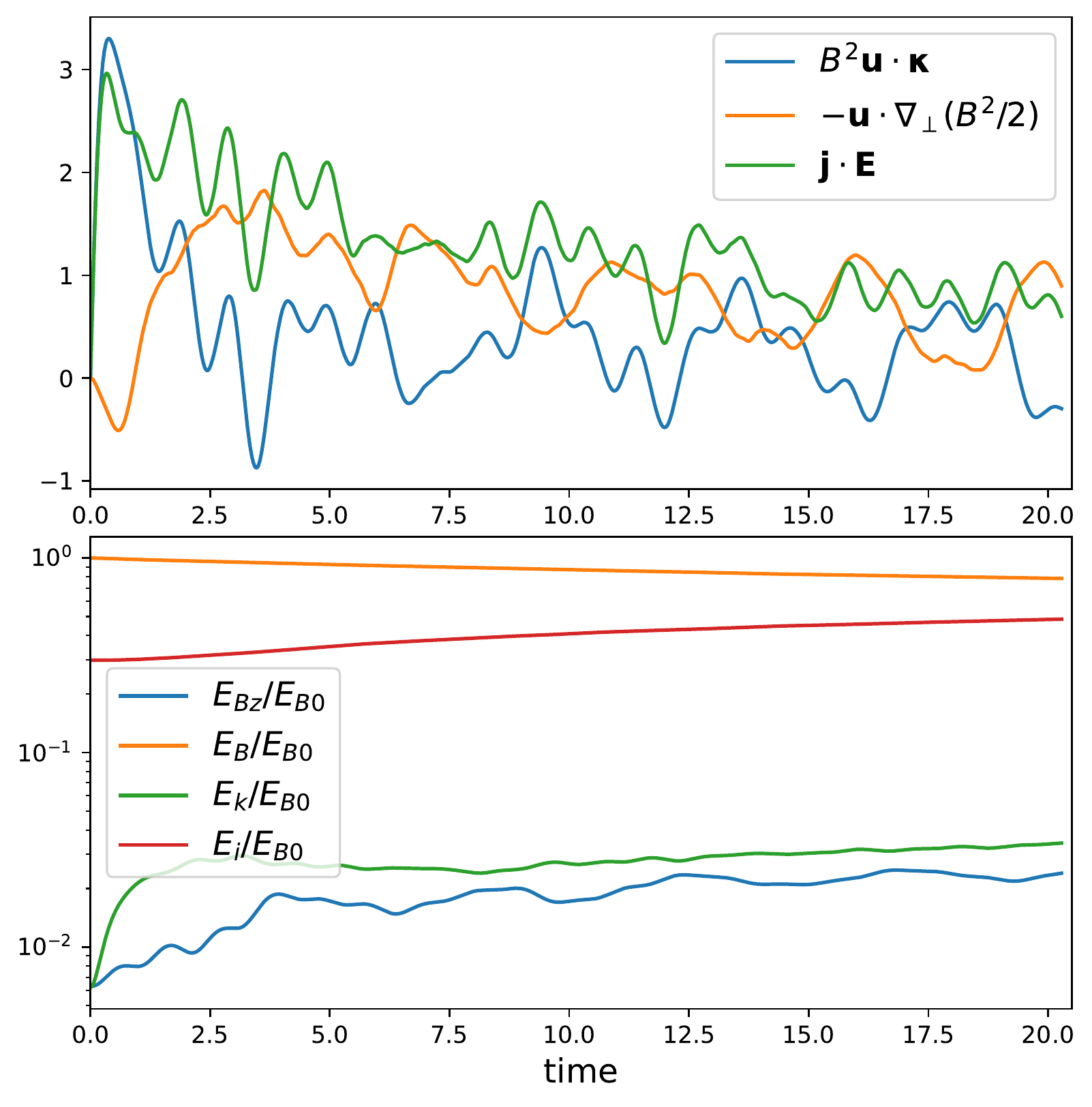}
  \caption{Top panel: time history of energization terms in Equation \eqref{eq:jdote2} for the magnetic reconnection simulation, including the curvature related term $(B^2/4\pi)\boldu\cdot\boldkappa$, the gradient related term $-\boldu\cdot\nabla_{\perp}(B^2/8\pi)$, and the total energization $\boldj\cdot\boldE$. All these terms are integrated over the entire simulation box. Bottom panel: evolution of different energy components as reconnection proceeds, including the $z$-component of magnetic field, total magnetic field, total kinetic energy, and internal energy. All the energies are normalized to the initial magnetic energy.}\label{fig:energy1}
\end{figure}

Figure \ref{fig:energy1} plots the energy conversion terms as in Equation \eqref{eq:jdote2} 
in the top panel. The evolution of energy in the $z$-component of the magnetic field 
 normalized to the initial magnetic energy is shown in the bottom panel,
which also includes the evolution of total magnetic energy, kinetic energy, and internal energy.
Based on our simulation, we find that the evolution may be separated into two stages. 
During the first stage, the evolution is characterized by a linear instability, which occurs 
at time $t \lesssim$ 4 as indicated by the initial exponential growth of energy in $B_z$ magnetic field.
For the second stage at later time, the growth rate slows down though 
the magnetic energy is still being released.
As illustrated by the top panel of Figure \ref{fig:energy1}, the CR term appears to be dominant
in energy conversion at the beginning of the first stage ($t \lesssim 2$).
At later time, however, the PE term becomes more important. The overall 
conversion of magnetic field is mostly contributed by the PE term at the end of the simulation.

\begin{figure}[!htp]
  \centering
  \includegraphics[width=0.333\linewidth]{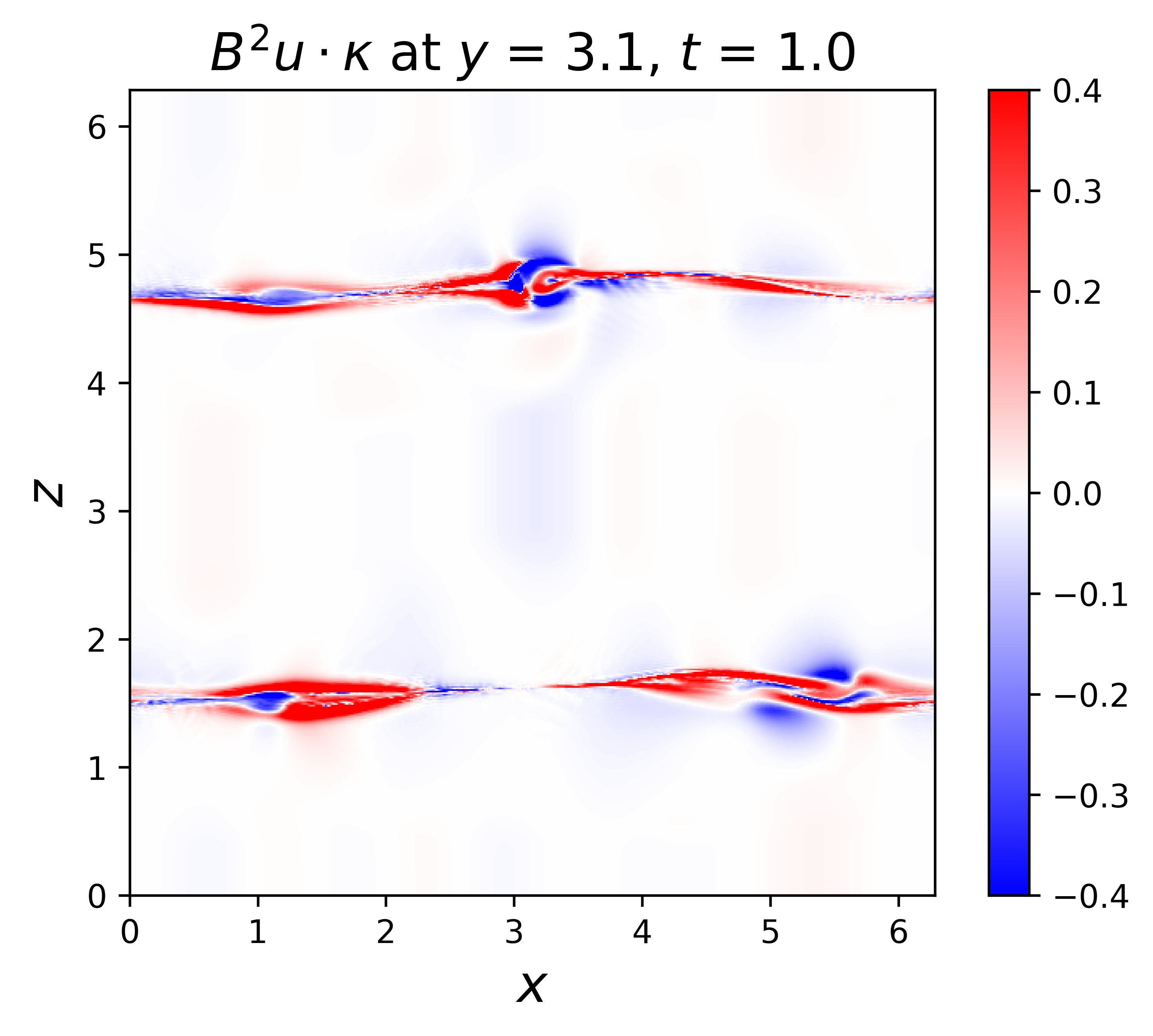}%
  \includegraphics[width=0.333\linewidth]{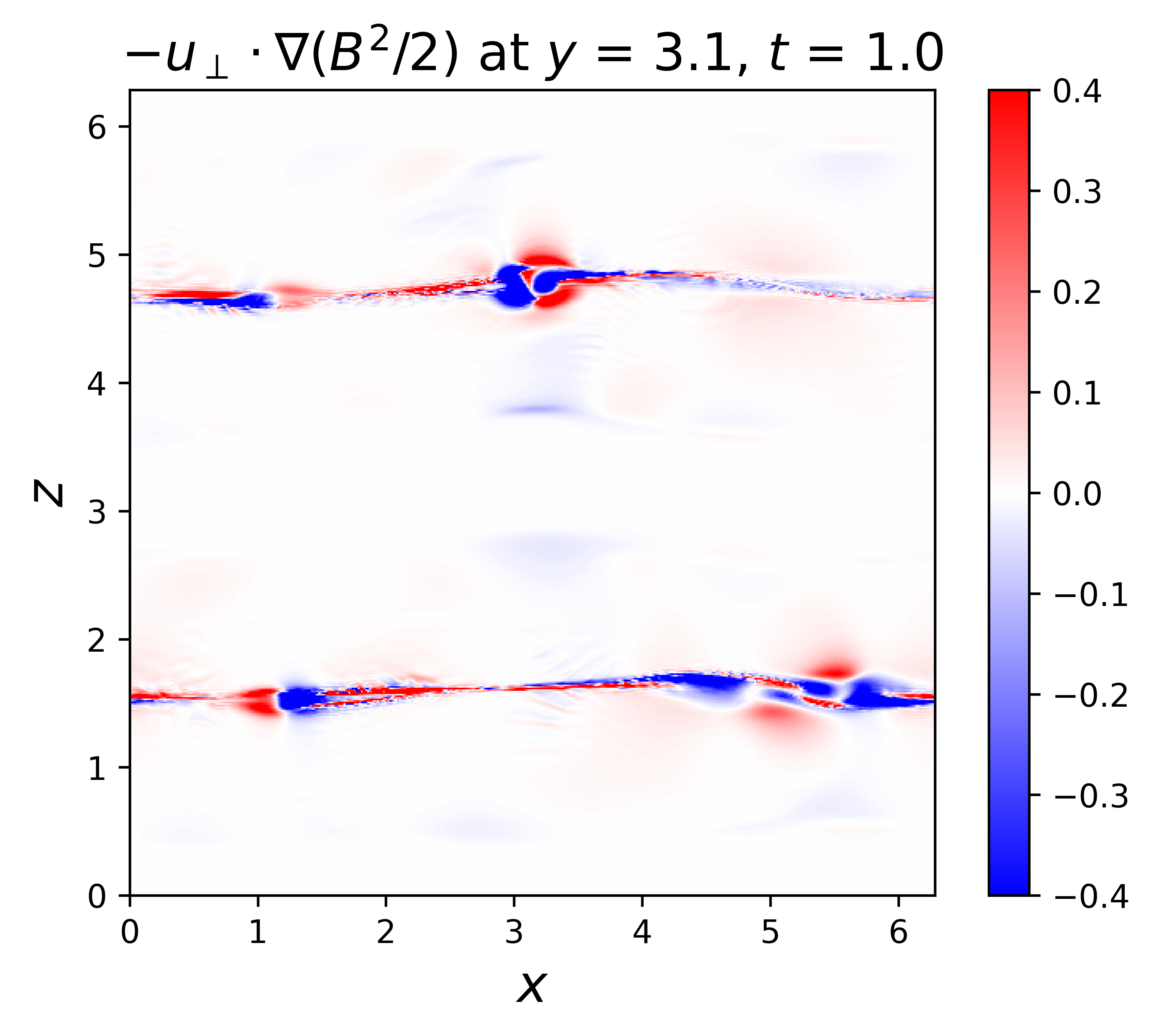}%
  \includegraphics[width=0.333\linewidth]{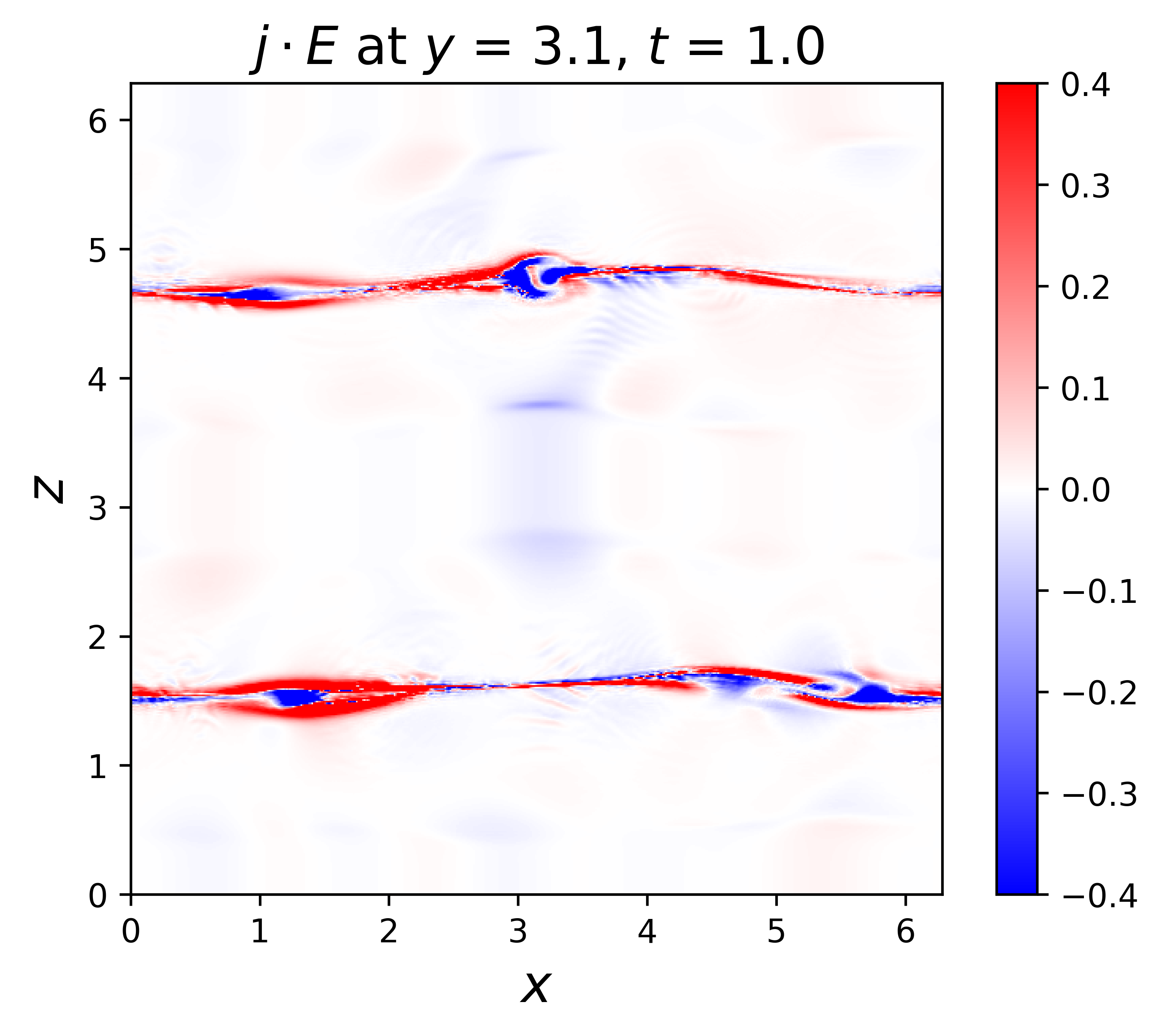}
  \includegraphics[width=0.333\linewidth]{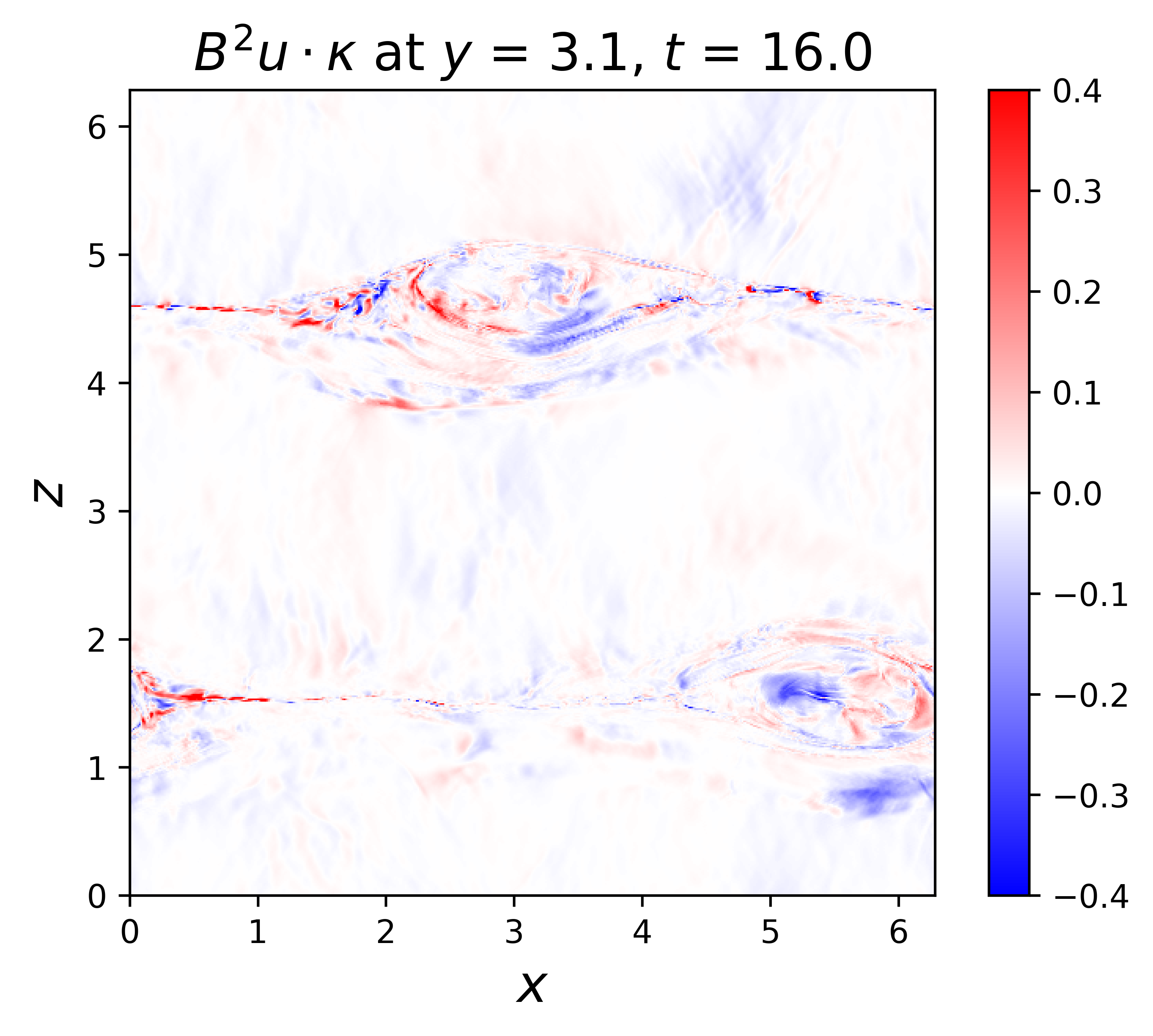}%
  \includegraphics[width=0.333\linewidth]{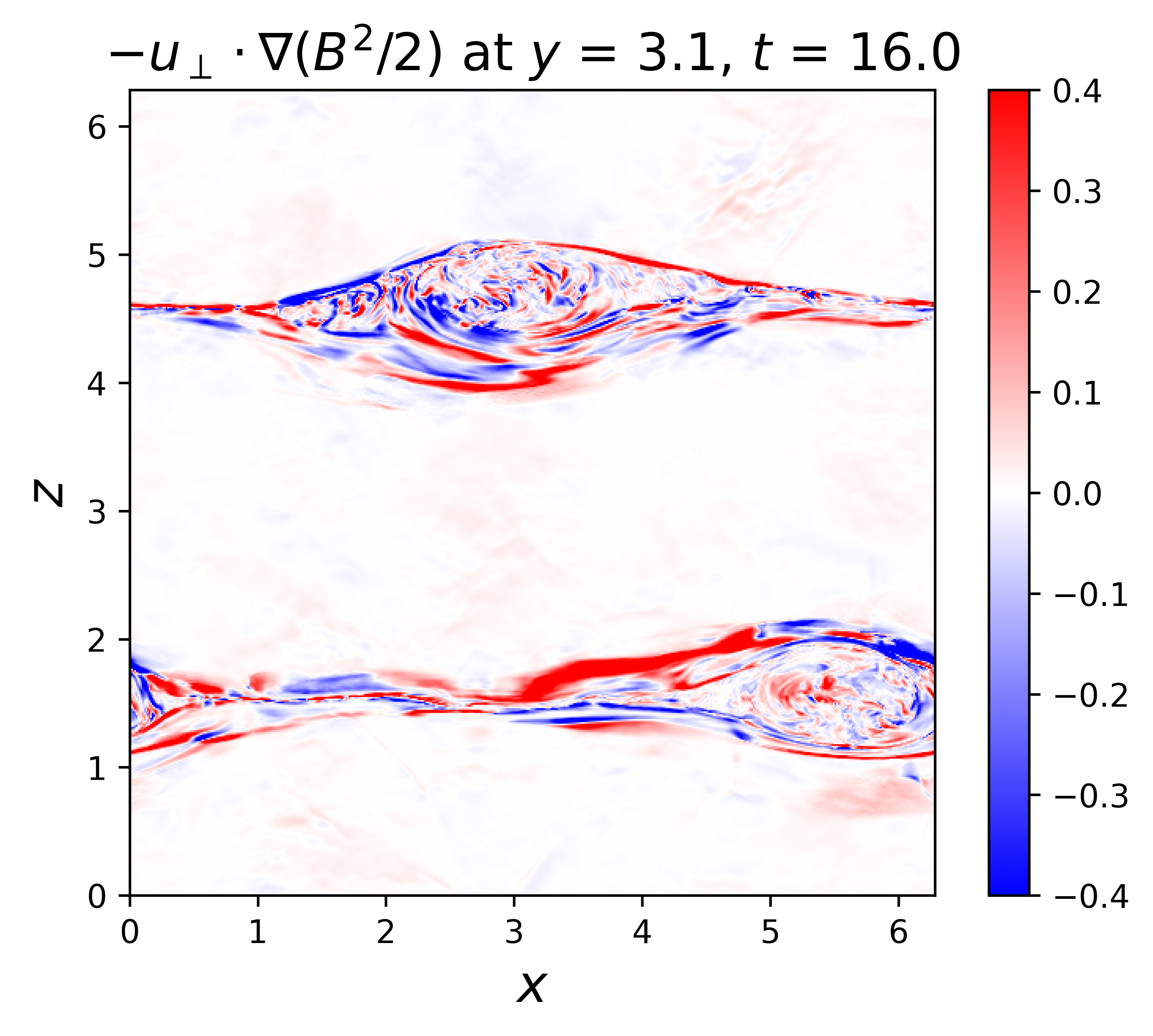}%
  \includegraphics[width=0.333\linewidth]{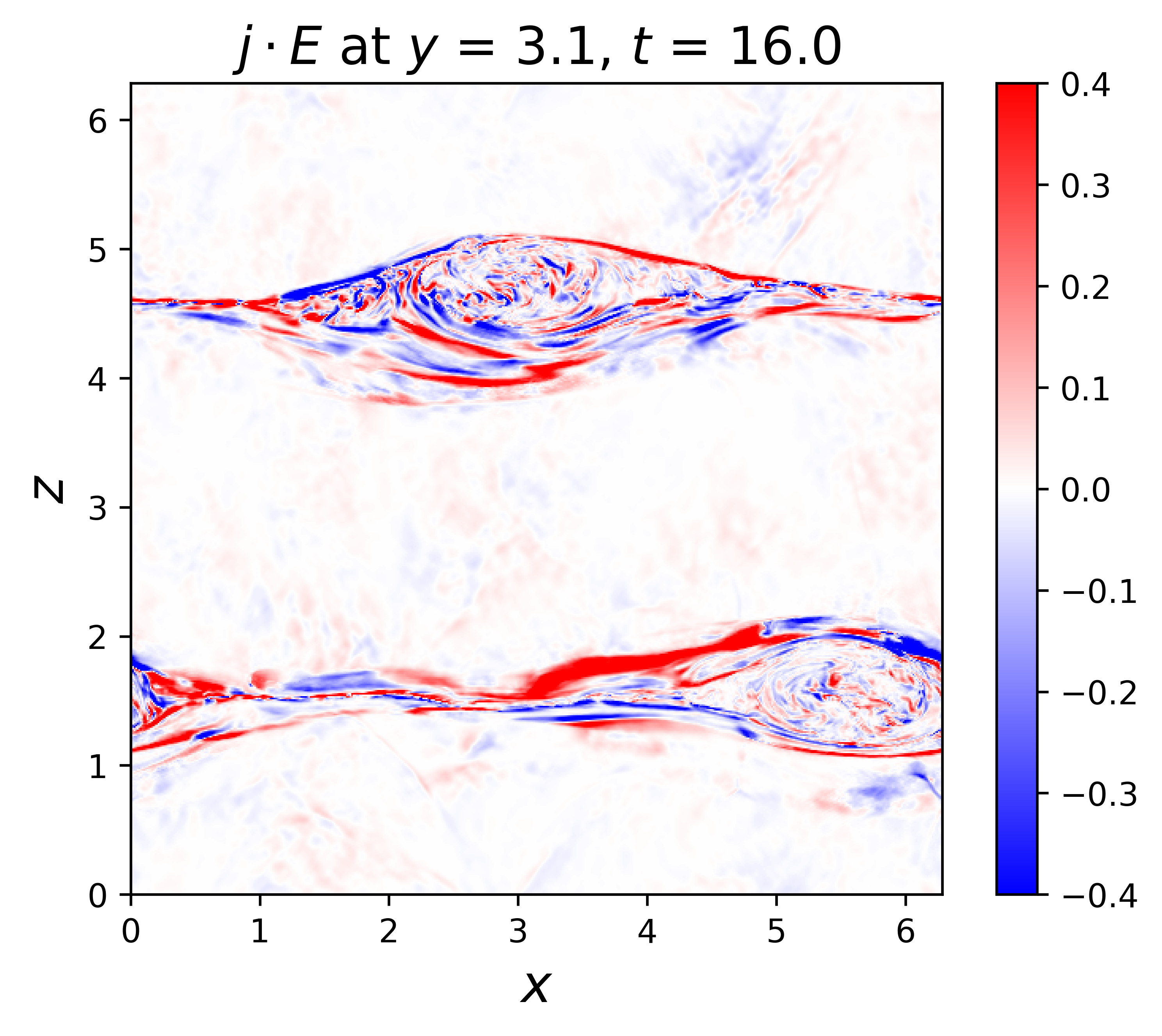}
  \caption{Spatial distribution of the CR (left) and PE (middle) terms 
  in the $x$-$z$ plane at early ($t = 1.0$) and late ($t = 16.0$) times of the 
  magnetic reconnection simulation. The right panels show the total energization
  as the sum of the CR and PE terms. }\label{fig:spatial1}
\end{figure}

Next, we look into the spatial distribution of the energy conversion terms. 
Figure \ref{fig:spatial1} displays the spatial distribution of the 
CR and PE terms and their sum in the $x$-$z$ plane.
Examples from the early ($t = 1.0$) and late ($t = 16.0$) stages are shown in the figure.
We find that during the first stage, the strong magnetic energy conversion 
due to relaxation of magnetic field curvature occurs near the reconnection 
X-line and the ends of magnetic islands formed in the two main current sheets.
It is interesting to note that there is an approximate spatial anti-correlation between
the CR and PE terms, although the CR term is stronger overall. 
During the second stage, both conversion terms appear to be turbulent, 
as there are small positive or negative patches
mixed in the reconnection and (3D) flux rope regions. The amplitude
of the PE term becomes much stronger and the global summation also
indicates its dominance. 

\begin{figure}[!ht]
  \centering
  \includegraphics[width=0.5\linewidth]{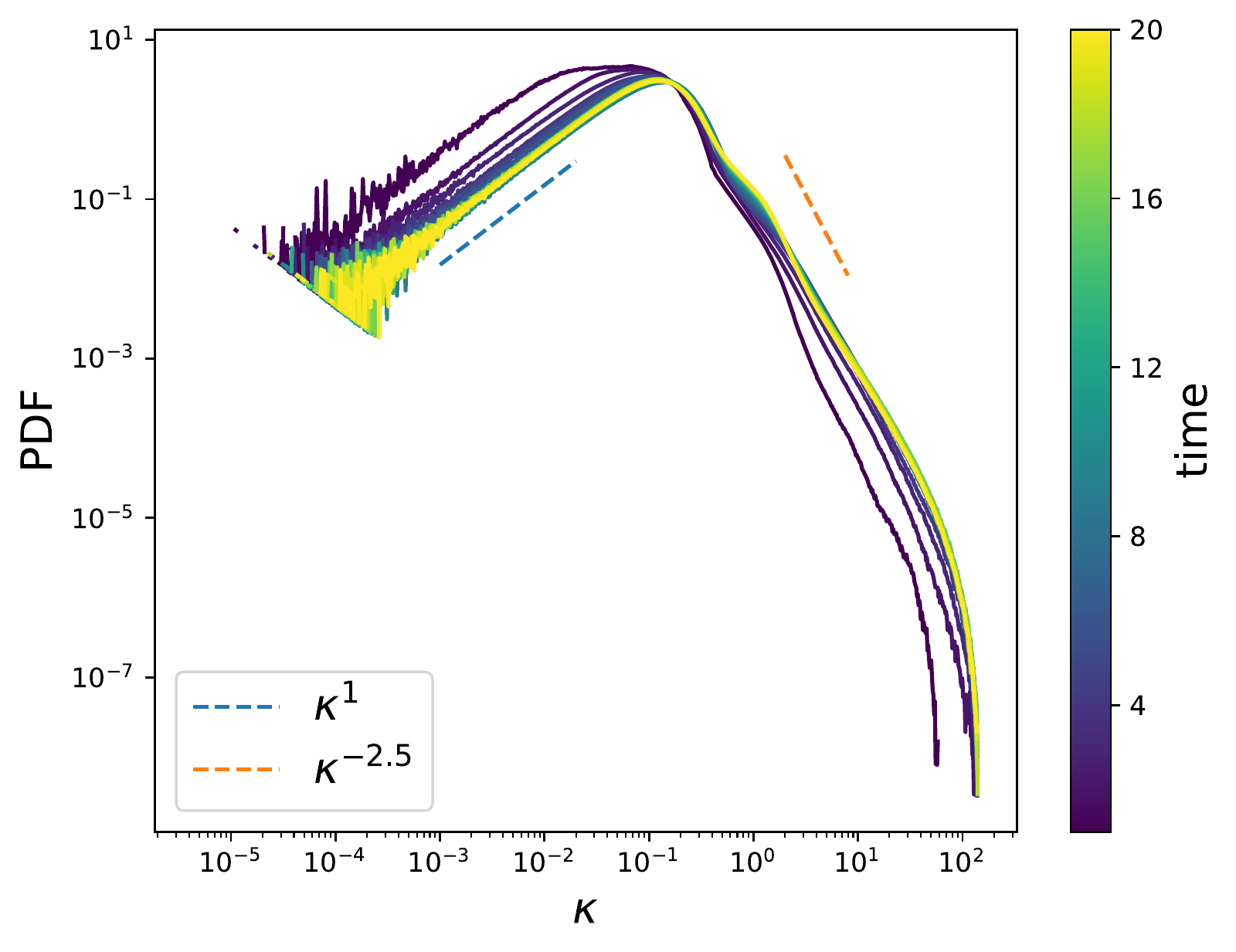}%
  \includegraphics[width=0.5\linewidth]{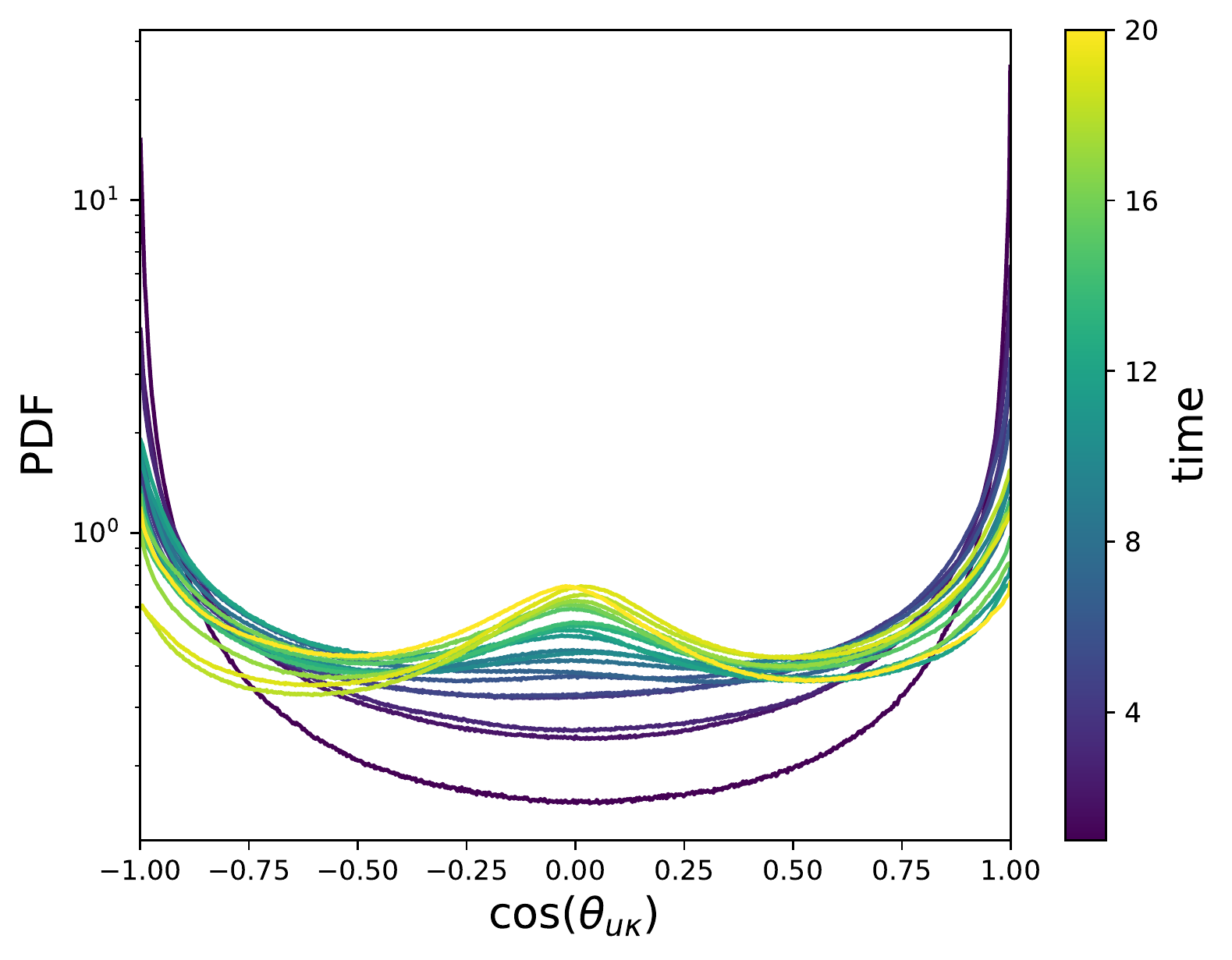}
  \caption{The probability density functions of $\kappa$ and $\cos\theta_{u\kappa}$ in the reconnection simulation. The colormap represent time evolution.}\label{fig:pdf1}
\end{figure}

To further illustrate the time evolution of the energy conversion terms, 
the probability density functions (pdfs) of the curvature magnitude 
$\kappa$ and the cosine of the angle between the velocity and 
curvature vector $\cos\theta_{u\kappa}$ are shown in Figure \ref{fig:pdf1}. 
The pdfs are calculated by the computational cell-based quantities.
The velocity distribution does not change much throughout the simulation 
and is not shown here. The curvature distribution extends to large 
curvature values and develops power law like distributions in the 
low and high curvature ranges, similar to those 
reported by \citet{Yang2019, Bandyopadhyay2020} 
in turbulence simulations and observations.
The pdfs of $\cos\theta_{u\kappa}$ show that at the early stage, 
the angle $\theta_{u\kappa}$ concentrates near $0^\circ$ and 
$180^\circ$ due to the reconnection inflow and outflow. 
The correlation between velocity and curvature directions 
degrades over time as the distribution becomes more 
random and tends toward zero. This is probably the reason 
why the magnetic energy conversion due to curvature relaxation becomes 
less dominant later in the simulation.

\begin{figure}[!ht]
  \centering
  \includegraphics[width=0.5\linewidth]{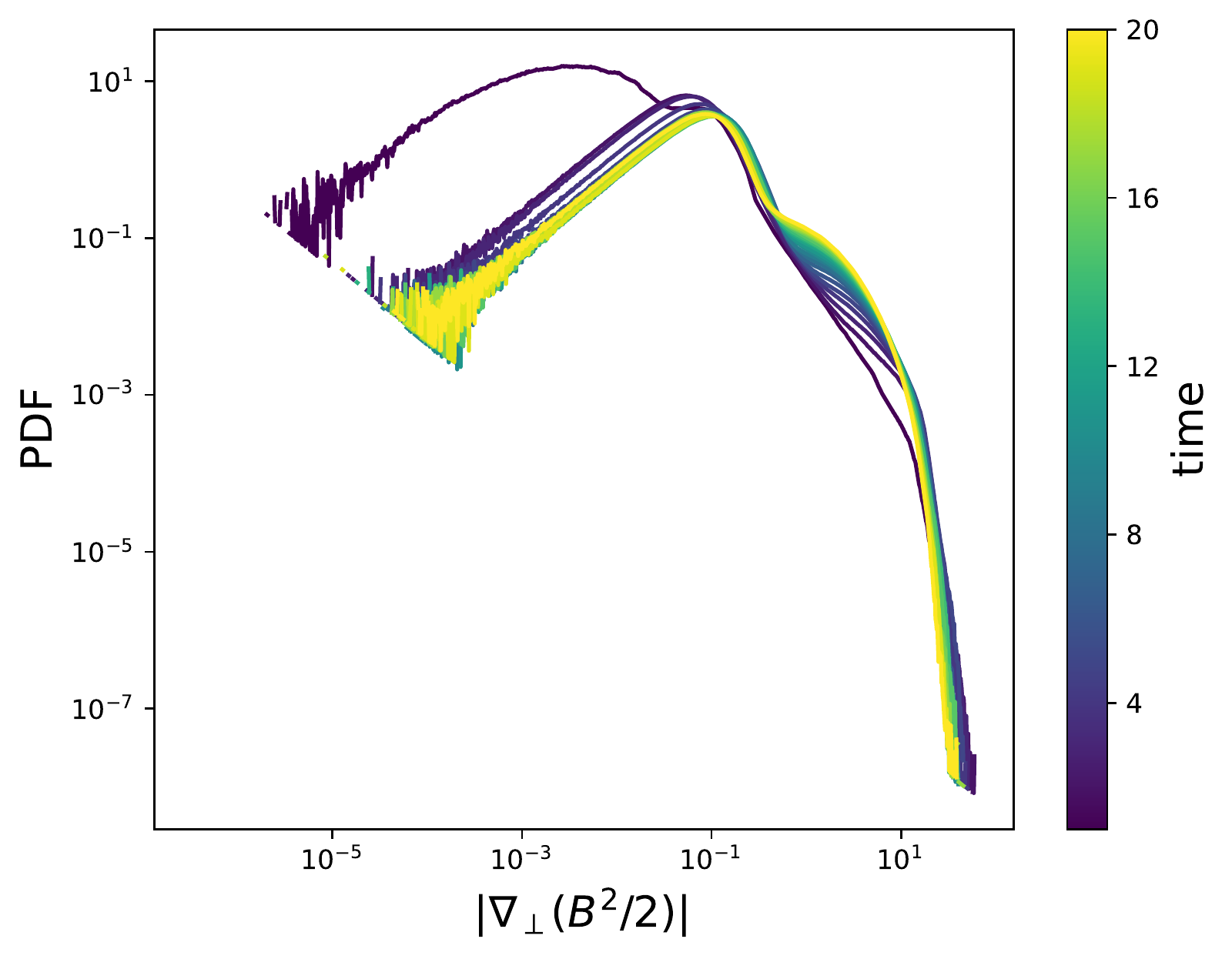}%
  \includegraphics[width=0.5\linewidth]{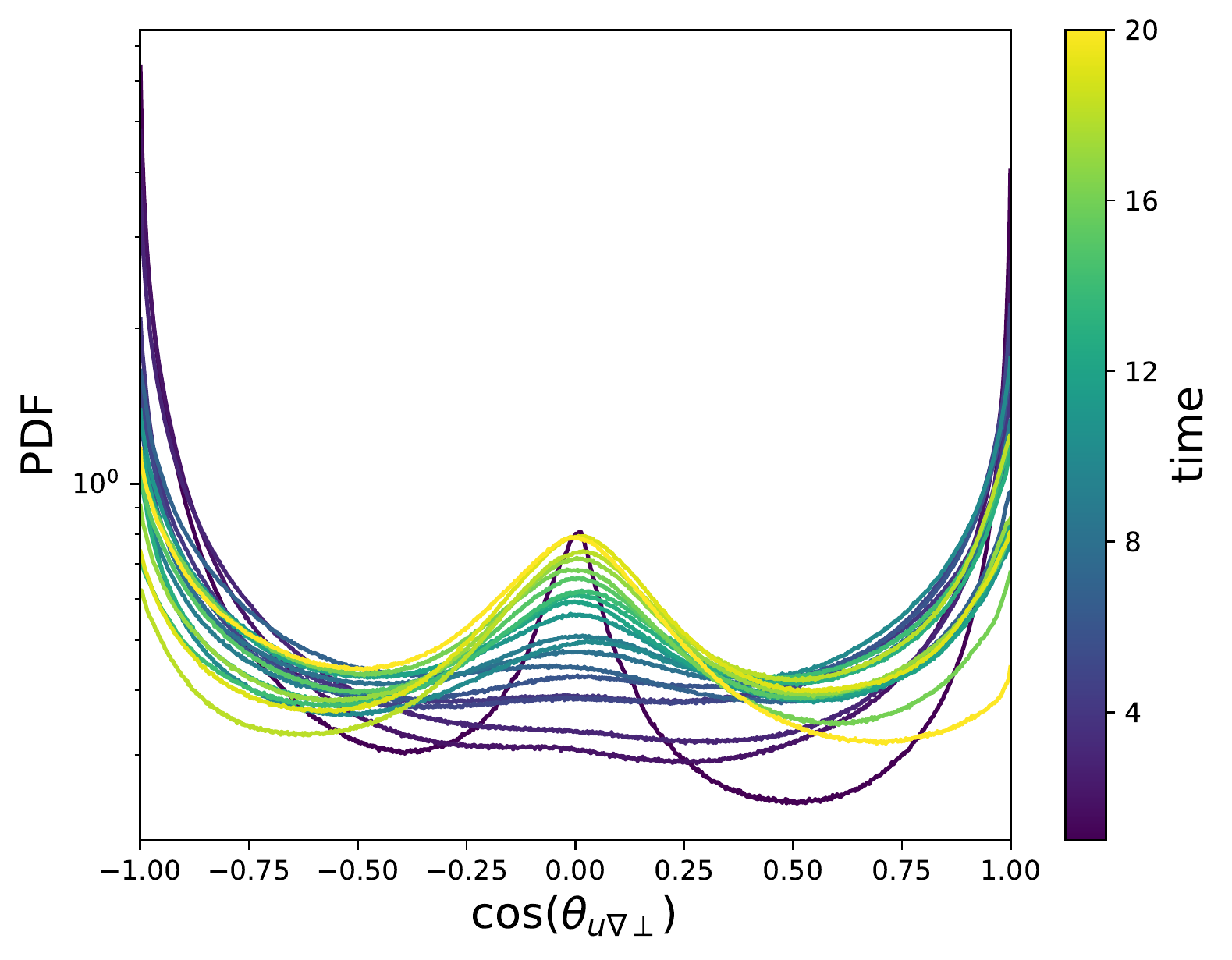}
  \caption{Similar to Figure \ref{fig:pdf1}, but for the pdfs of $|\nabla_{\perp}(B^2/2)|$ and $\cos\theta_{u\nabla\perp}$ (where $\theta_{u\nabla\perp}$ is the angle between $\boldu$ and $\nabla_{\perp}(B^2/2)$) in the reconnection simulation. The colormap represent time evolution. }\label{fig:pdf2}
\end{figure}

Similarly, we can also inspect the pdfs of the perpendicular 
gradient $\nabla_{\perp}(B^2/2)$ and the angle it 
makes with the velocity $\theta_{u\nabla\perp}$, 
as shown in Figure \ref{fig:pdf2}.
The correlation between the velocity and perpendicular gradient 
vectors also degrades over time.
Unlike the curvature distribution, however,
the pdf of $|\nabla_{\perp}(B^2/2)|$ 
develops a ``bump" starting at high values $\sim 0.5$, which leads to the 
dominance of the perpendicular expansion process. 
This bump comes from the turbulent patches as shown
in the lower right panel of Figure \ref{fig:spatial1}. 
 
\begin{figure}[!ht]
  \centering
  \includegraphics[width=0.5\linewidth]{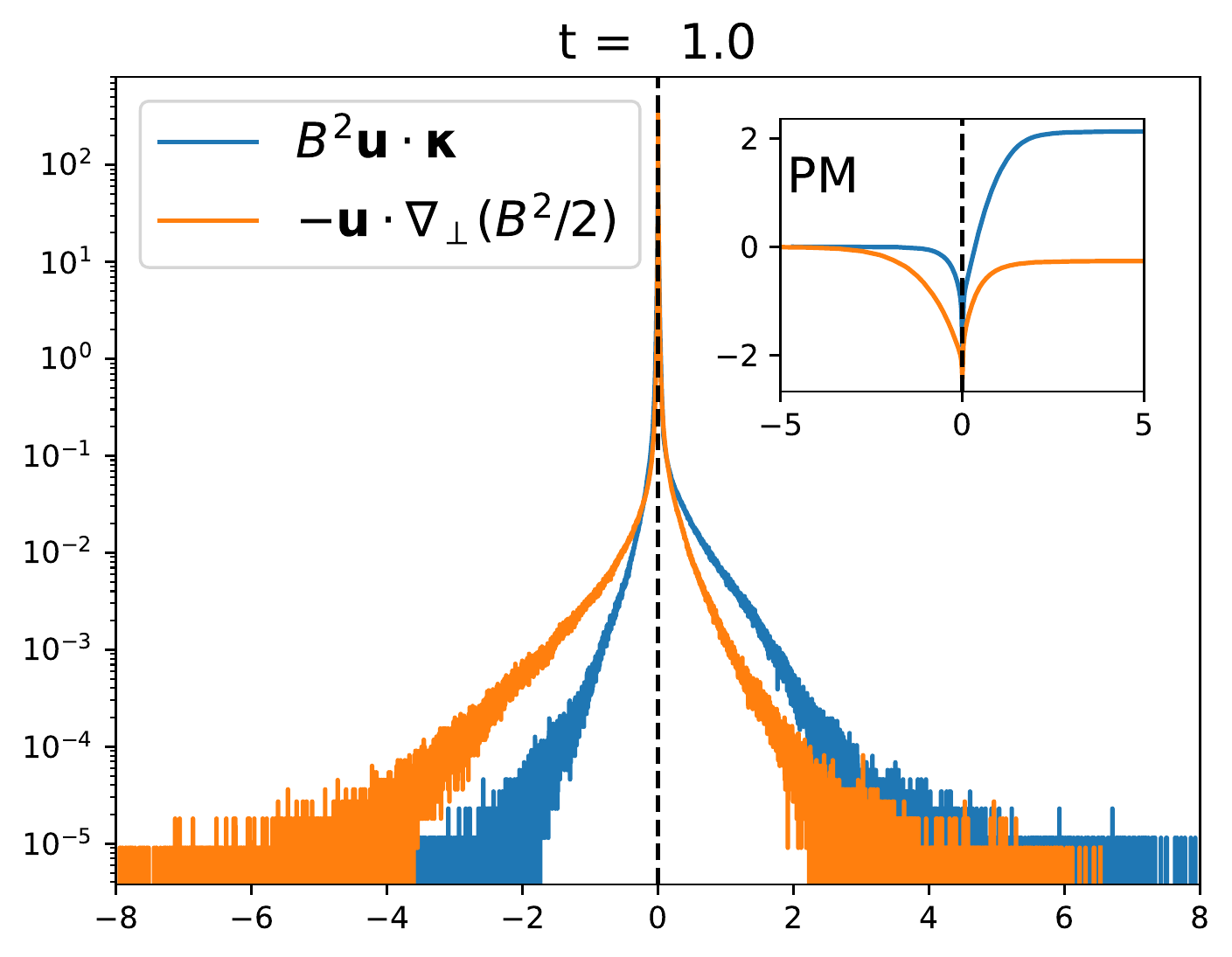}%
  \includegraphics[width=0.5\linewidth]{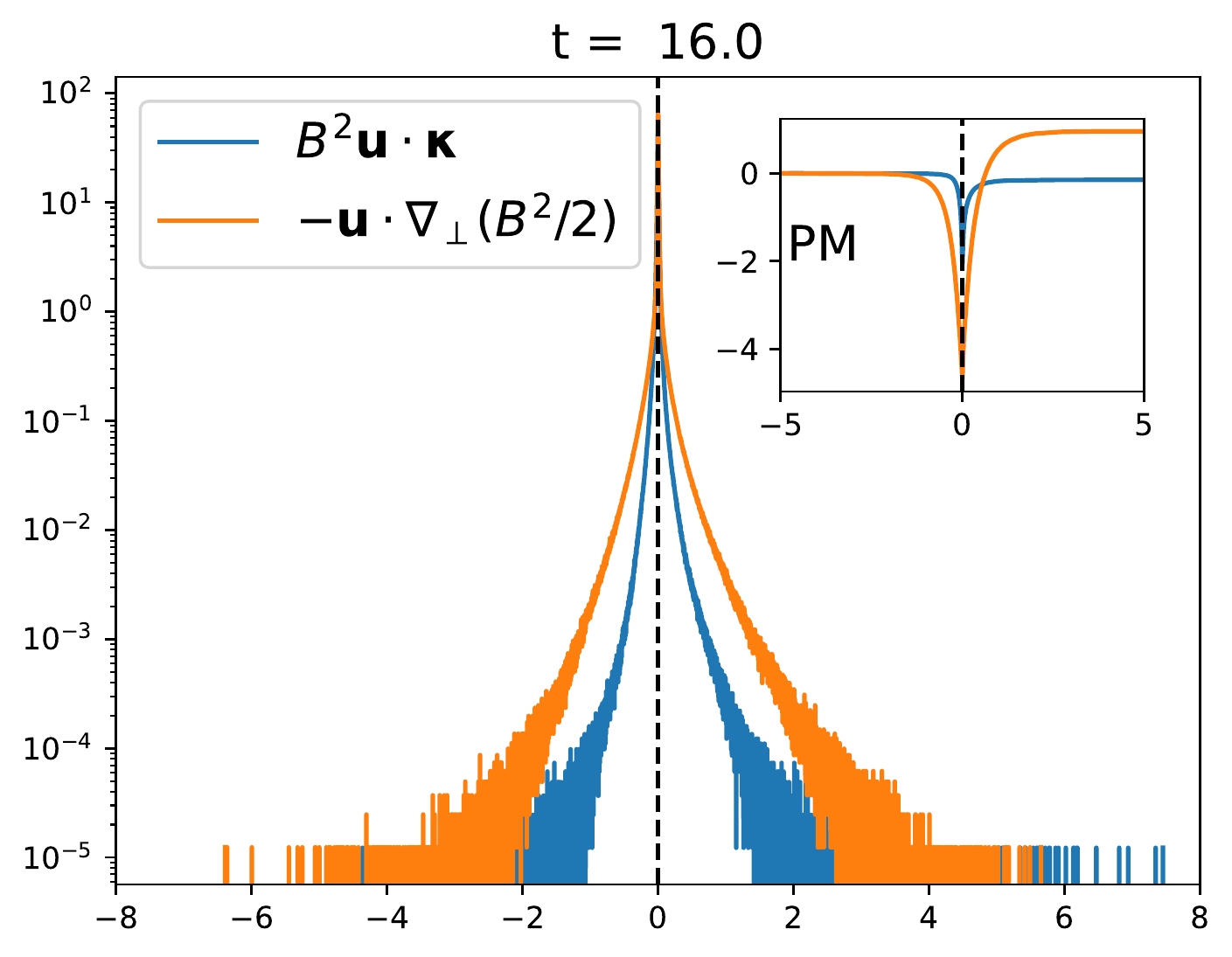}
  \caption{The probability density functions of the two energization terms $B^2 \boldu\cdot\boldkappa$ and $-\boldu\cdot\nabla_{\perp}(B^2/2)$ in the reconnection simulation at early and late times. The insets show the cumulative partial moments (PM) from the two terms (see text for the definition).}\label{fig:pdf3}
\end{figure}

Figure \ref{fig:pdf3} shows the pdfs of the two conversion terms 
in Equation \eqref{eq:jdote2} at two different time frames.
At an earlier time ($t = 1$), the distributions of both terms are 
clearly skewed, suggesting systematic effects of the positive 
velocity-curvature correlation and perpendicular fluid compression. 
In contrast, at a later time ($t = 16$), both pdfs appear to be 
more symmetric about zero due to the more turbulent nature of the system.
The insets show the cumulative partial moments $PM(x) = \int_{-\infty}^{x} x'f(x') dx'$
with $f(x')$ being the pdf of $B^2 \boldu\cdot\boldkappa$ or $-\boldu\cdot\nabla_{\perp}(B^2/2)$.
Although most of the cells have nearly zero contribution to the 
energization as indicated by the sharp peaks in the pdfs, 
the partial moments illustrate that the cells with $-\boldu\cdot\nabla_{\perp}(B^2/2) \geq 0.5$ contribute 
significantly to the total magnetic energy conversion.
These analyses strongly suggest that the overall magnetic energy 
conversion in 3D reconnection is actually dominated by the development
of turbulent patches inside the flux ropes undergoing 
perpendicular expansion owing to the magnetic pressure gradients.
These patches could be consequences of collisions between
reconnection outflows as well as secondary instabilities as discussed
in \citet{HuangY2016, Kowal2017, YangL2020}.

\begin{figure}[!htp]
  \centering
  \includegraphics[width=0.5\linewidth]{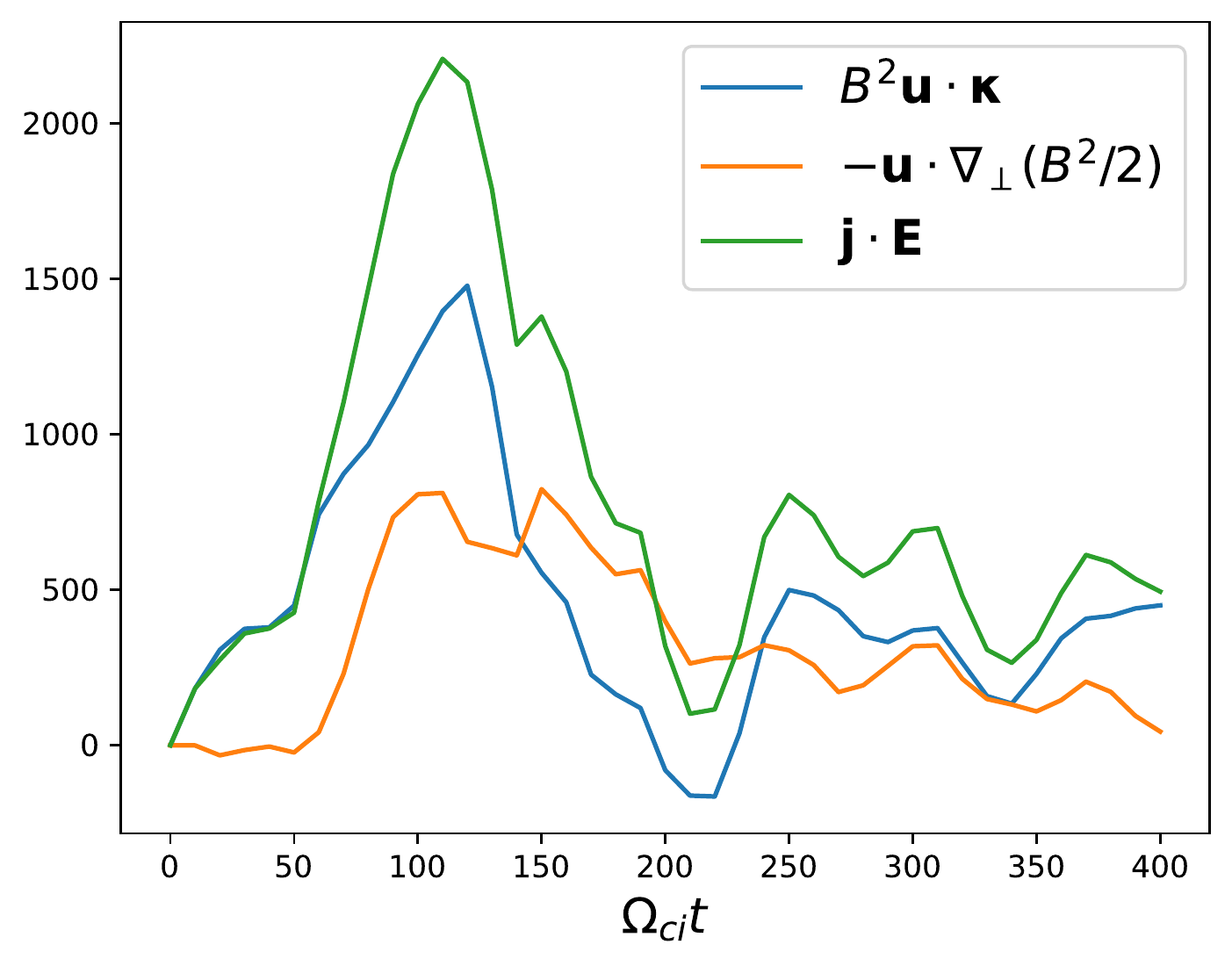}
  \caption{Decomposition of the $\boldj\cdot\boldE$ term using the 3D PIC simulation 
  by \cite{Li2019} which has a guide field of $B_g = 0.2$. This guide field weakens
  the effect of the PE process, making the CR process more dominant.}\label{fig:energy-PIC3d}
\end{figure}

\begin{figure}[!htp]
  \centering
  \includegraphics[width=0.5\linewidth]{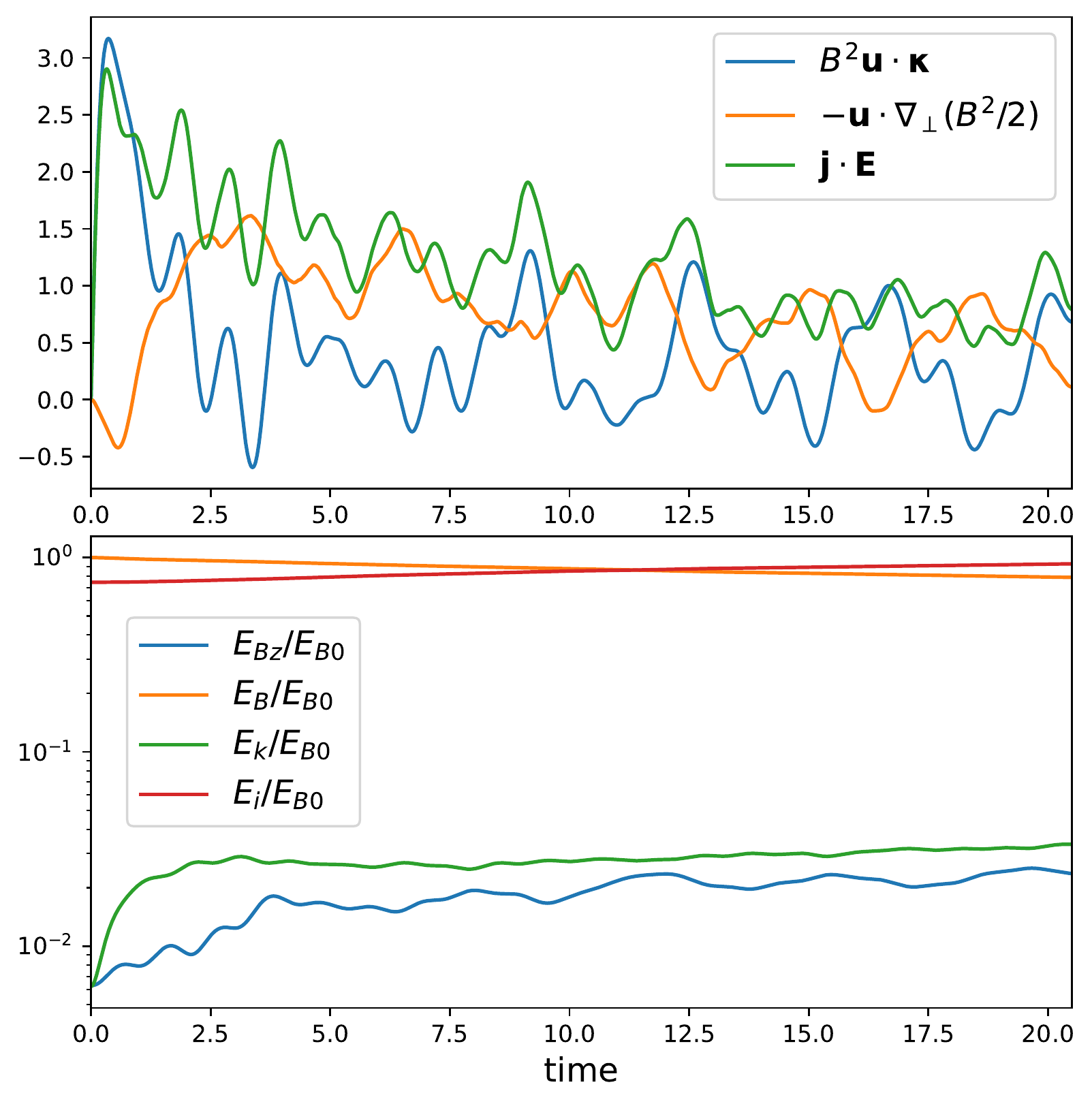}%
  \includegraphics[width=0.5\linewidth]{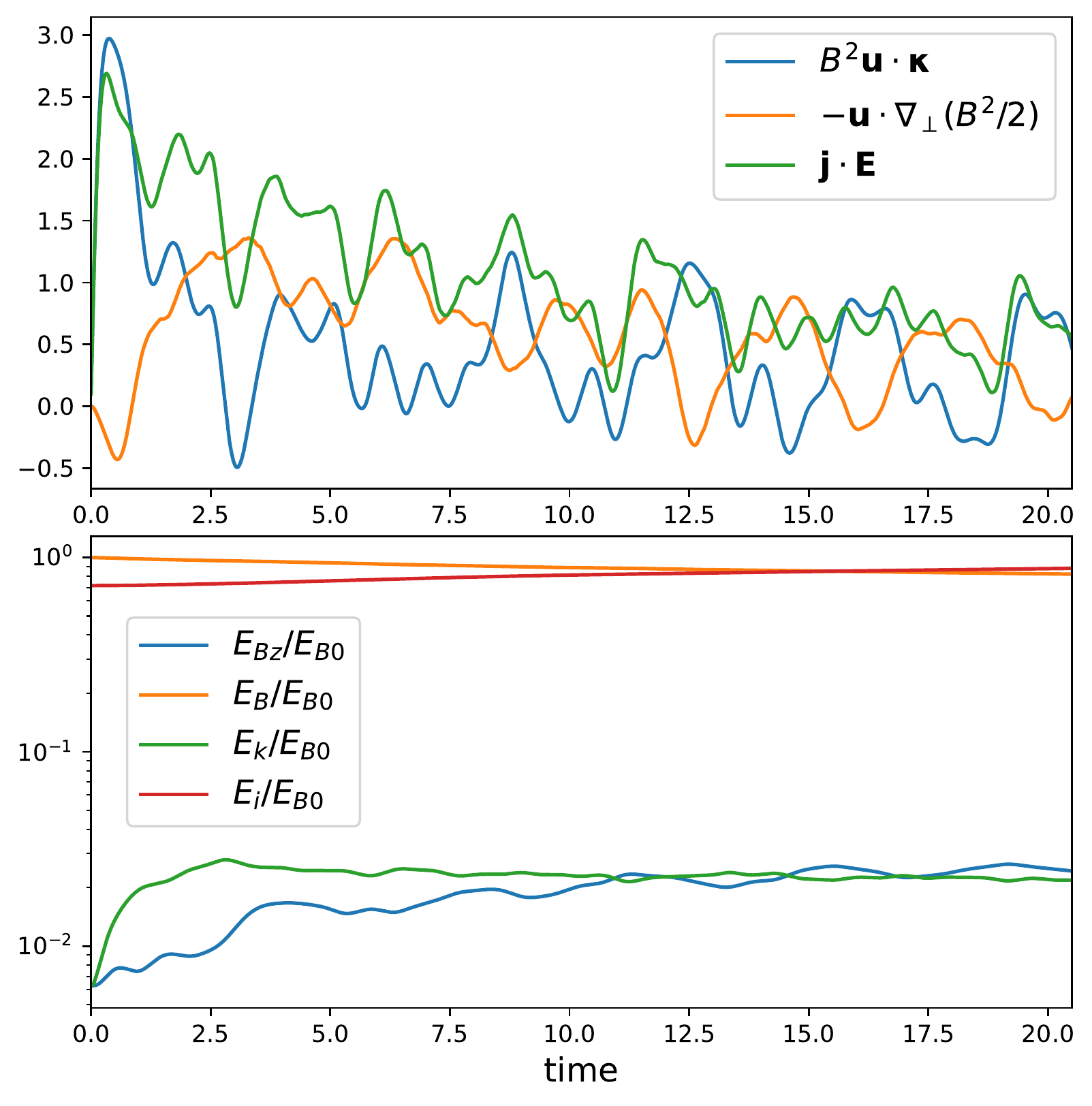}
  \caption{History of the energization terms similar to Figure \ref{fig:energy1}, but for $\beta=0.5$.
  The left panel is for the case without guide field and the right panel has a guide field of $B_g = 0.2 B_0$.}\label{fig:energy-b05}
\end{figure}

\begin{figure}[!htp]
  \centering
  \includegraphics[width=0.5\linewidth]{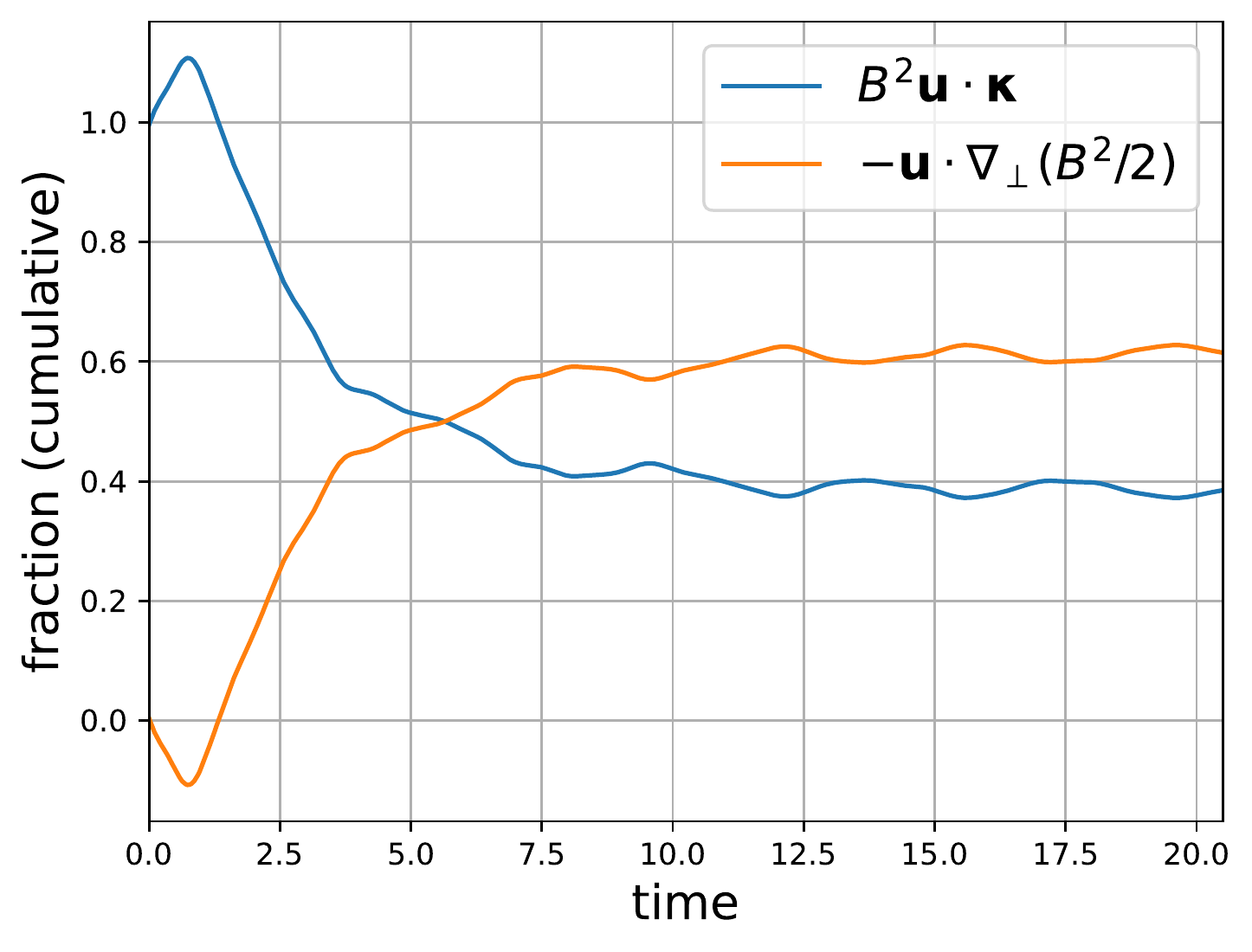}%
  \includegraphics[width=0.5\linewidth]{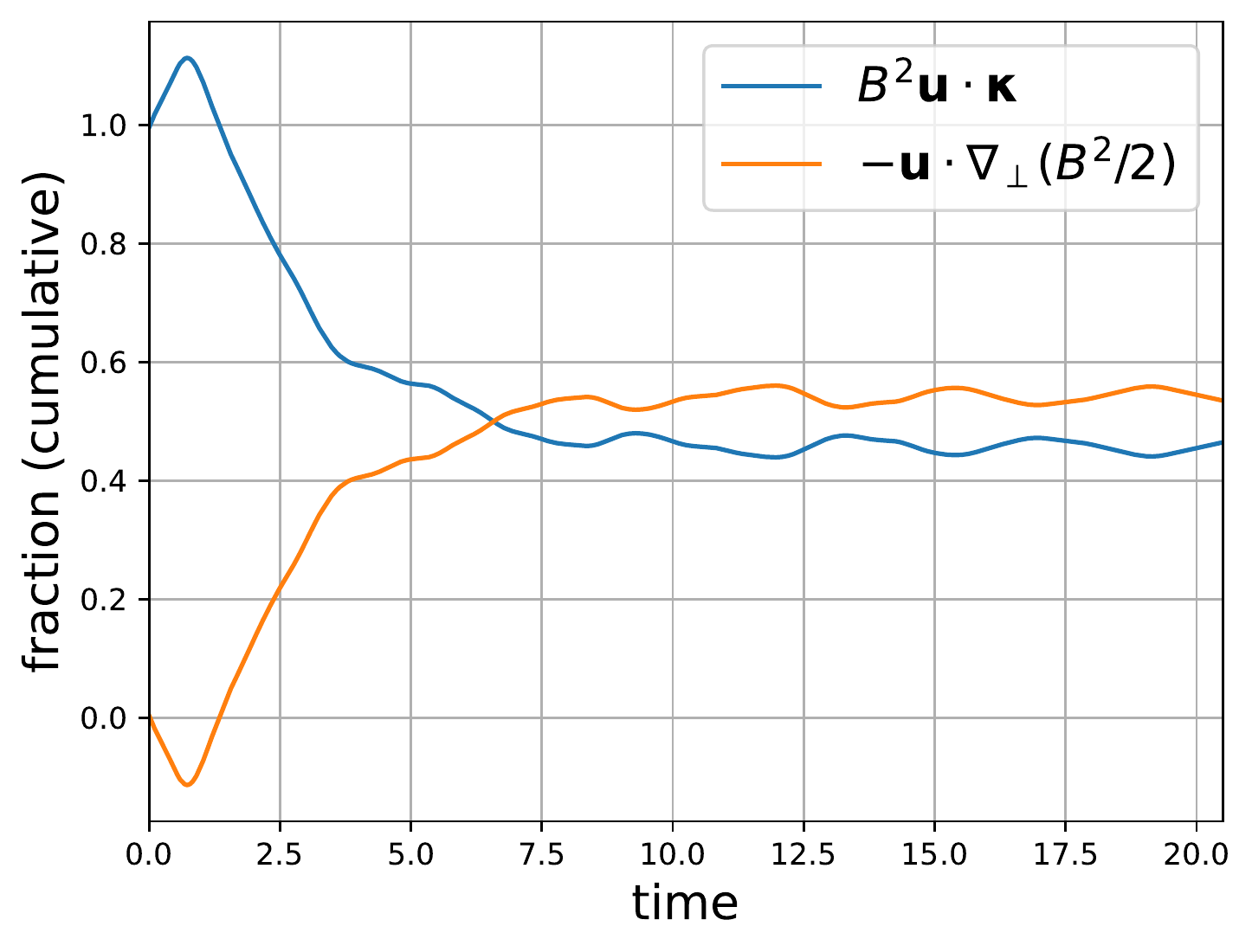}
  \caption{Fractions of the energization terms in total $\boldj\cdot\boldE$ for $\beta=0.5$.
  The left panel is for the case without guide field and the right panel has a guide field of $B_g = 0.2 B_0$.}\label{fig:frac-b05}
\end{figure}

To further investigate how the CR and PE terms behave under different conditions,
we have studied a 3D reconnection set-up with a guide field $B_g = 0.2$ using
Equation (\eqref{eq:rec_init}. This is the same set-up as given
in a 3D particle-in-cell (PIC) simulation presented by \cite{Li2019}.
The PIC simulation has a similar initial magnetic field configuration as our MHD simulation,
though with only one current sheet, and the box size is $150 d_i \times 75 d_i \times 62.5 d_i$
where $d_i$ is the ion inertial length.
Figure \ref{fig:energy-PIC3d} shows the evolution of both CR and PE terms,
though interestingly the curvature relaxation term appears to be slightly 
more important overall. This is different from the case when the guide field  $B_g = 0$.
We note that the normalization here is different from the MHD simulation.
The velocity is normalized to the speed of light and length to the electron inertial length $d_e$.
The electron plasma frequency $\omega_{pe}$ is set equal to the electron cyclotron frequency $\Omega_{ce}$,
which is defined by the magnetic field strength in the reconnection ($x$-$z$) plane.
The ion-to-electron mass ratio is set to 25 so that the Alfv\'en speed is $V_A \simeq d_i \Omega_{ci} = 0.2 c$.
We have verified that including the same guide field in the 3D MHD reconnection 
simulation will also enhance the relative importance of the curvature relaxation term. 
The physical reason may be that a guide field component can suppress the 
expansion of magnetic field line perpendicular to the upstream magnetic field.
Figure \ref{fig:energy-b05} shows the history of the energization terms
for two runs with $\beta=0.5$. The format is the same as Figure \ref{fig:energy1}.
The case without guide field is shown in the left panel and the case with a guide field
$B_g = 0.2 B_0$ is shown in the right panel. The general evolution of these cases
are similar to the $\beta=0.2$ run. Figure \ref{fig:frac-b05} plots the fraction
of the CR and PE contributions to the total energization (cumulative in time),
which illustrates more clearly that the inclusion of a guide field enhances
the relative importance of the curvature relaxation term.

\subsection{Kink jet}

Another common astrophysical system that may lead to strong 
magnetic energy conversion is kink instability in astrophysical jets. 
Here, we present a relativistic MHD simulation of a kink unstable jet
following the setup in \citet{Mizuno2009} and \citet{Zhang2017}.
The magnetic field has the form of
\[ B_z = \frac{B_0}{1 + (r/r_0)^2}; B_\phi = \frac{B_0}{1 + (r/r_0)^2} \kl{\frac{r}{r_0}}, \]
where we use $B_0=2$ and $r_0=1$; $r$ is the 
radial distance to the central axis.
The simulation box size is 40 in $x$ and $y$ directions 
with 640 cells each and 64 in $z$ direction with 1024 cells.
Periodic boundaries are used in $z$ direction and outflow 
boundaries in $x$ and $y$ directions.
The system is perturbed by random velocity fluctuations 
of 0.01 $c$ ($c = 1$ in the simulation).
The simulation is run until the time of 350.
The initial density $\rho$ and pressure $p$ are uniform ($\rho = 1$ and $p = 0.01$),
and plasma $\beta$ at the central axis is $5\times10^{-3}$.
The magnetization parameter $\sigma = E_{em}/h$ is about 2 at the central axis,
where $E_{em}$ is the electromagnetic energy density, and $h = \rho c^2 + \gamma p / (\gamma-1)$
is the specific enthalpy with $\gamma = 4/3$ the adiabatic index.

\begin{figure}[!htp]
  \centering
  \includegraphics[width=0.5\linewidth]{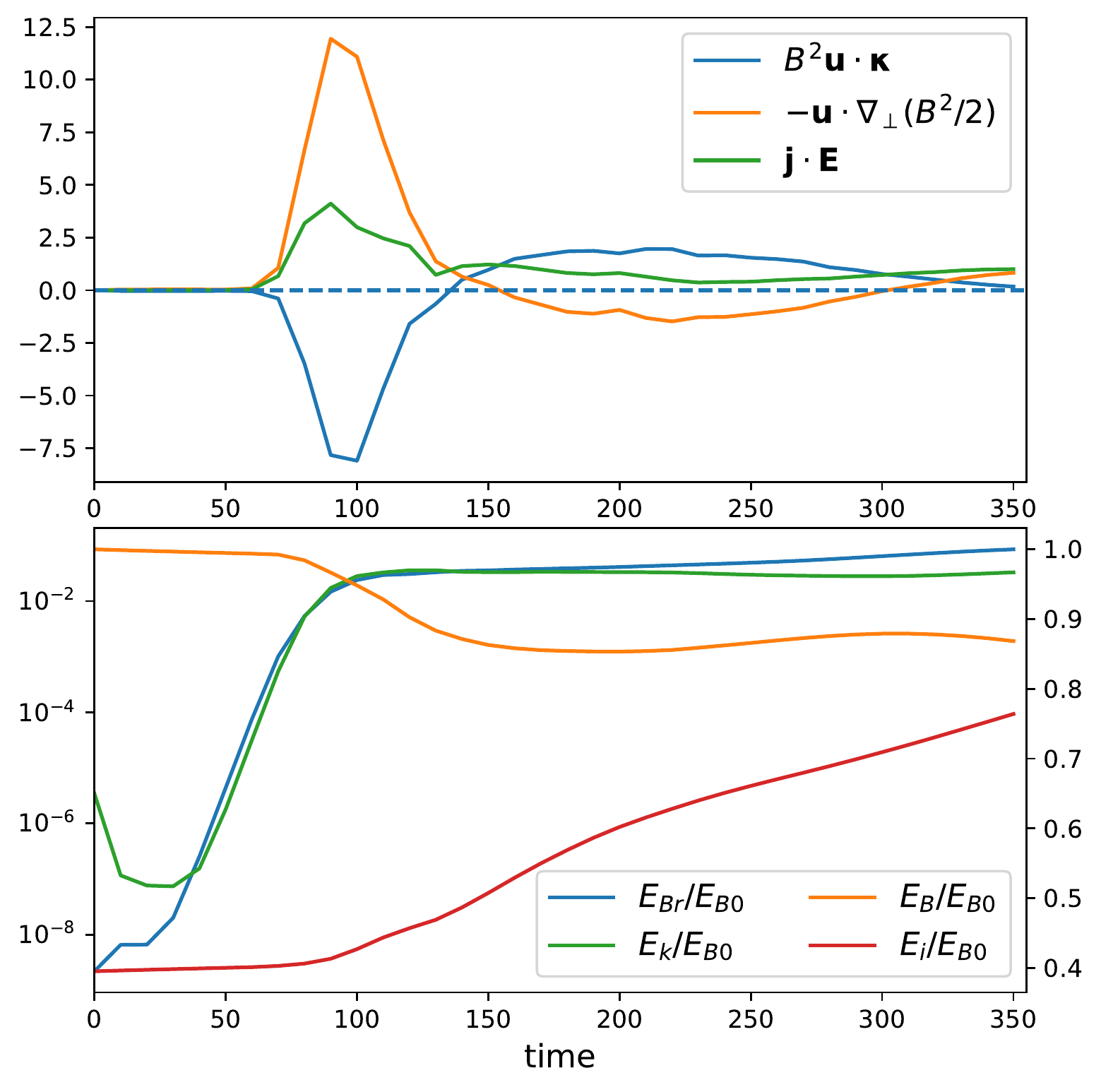}
  \caption{Time history of the energization terms and the energy components similar to Figure \ref{fig:energy1}, but for the kink simulation. The total magnetic energy and the internal energy are plotted in linear scale, and the energy in $B_r$ and total kinetic energy in log scale.
  }\label{fig:energy2}
\end{figure}

We apply the same analysis as in the reconnection simulation. 
Similar to Figure \ref{fig:energy1}, we plot the time history of the 
magnetic energy conversion terms in the top panel of Figure \ref{fig:energy2}. 
The bottom panel of Figure \ref{fig:energy2} plots the evolution of energy 
in the radial component of magnetic field developed from the kink instability. 
The total magnetic energy, kinetic energy, and internal energy are also plotted in the figure.
To avoid the effect of the open boundaries, we restrict the calculation to the region
$|x| \le 10$ and $|y| \le 10$. At later time $t \gtrsim 200$ (which is 5 light-crossing times in the transverse direction), the total magnetic energy stops decreasing
while the total magnetic energy conversion $\boldj\cdot\boldE$ is positive, which is likely a boundary effect.
Nevertheless, similar to the reconnection case, the system undergoes a rapid instability 
in the beginning, as illustrated by the exponential increase 
in the $E_{Br}$ energy before $t \sim 100$.
At the later stage of the evolution, the instability saturates and the 
system enters a more turbulent state.
The top panel of Figure \ref{fig:energy2} shows that the 
total energization ($\boldj\cdot\boldE$) is nearly zero in the earlier stage 
and maintains a finite positive value at the later stage.
It can also be seen that the CR and PE terms are anti-correlated
very well with each other and tend to cancel each other out.
This can be understood as both the curvature and gradient vectors 
of the magnetic field point toward the central axis in a cylindrical 
kink configuration at zeroth order, so that the two terms 
in Equation \eqref{eq:jdote2} have opposite signs.
And it can indeed be verified that $\boldkappa = \nabla_{\perp}(B^2/2) / B^2$
for the initial kink magnetic field configuration.
The finite energization in the later stage is mostly 
contributed by the CR term, though part of its contribution 
is canceled out by the PE term.
As the kink instability becomes
further developed, the $\boldj\cdot\boldE$ term shows that the net magnetic 
energy conversion becomes appreciable after $t \sim 80$ when the 
PE term becomes the main contributor in the magnetic energy conversion
though part of its contribution is canceled out by the CR term.
At the later more turbulent stage, the roles of the two terms are reversed
and the net magnetic energy conversion rate is reduced.

\begin{figure}[!htp]
  \centering
  \includegraphics[width=0.33\linewidth]{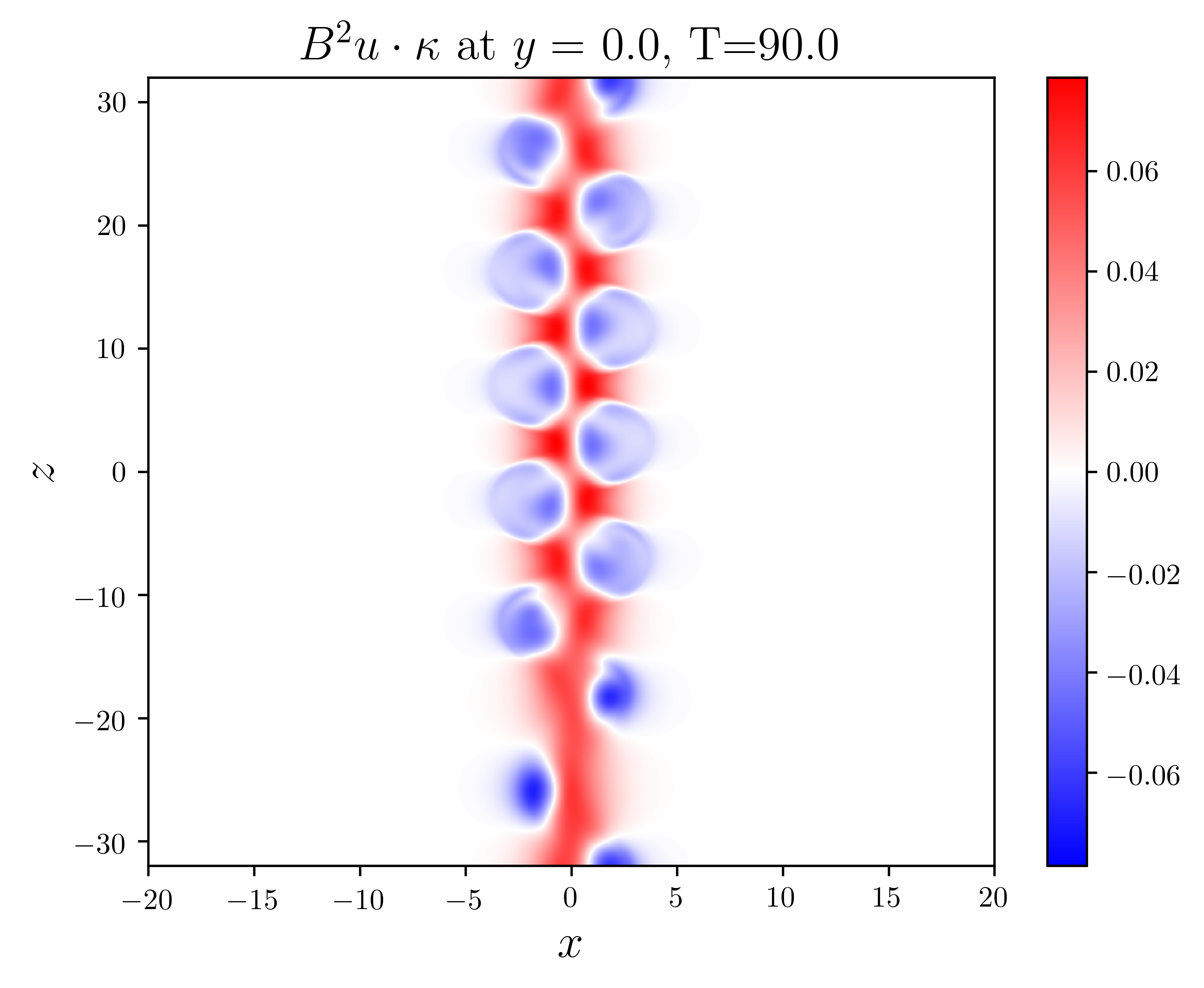}%
  \includegraphics[width=0.33\linewidth]{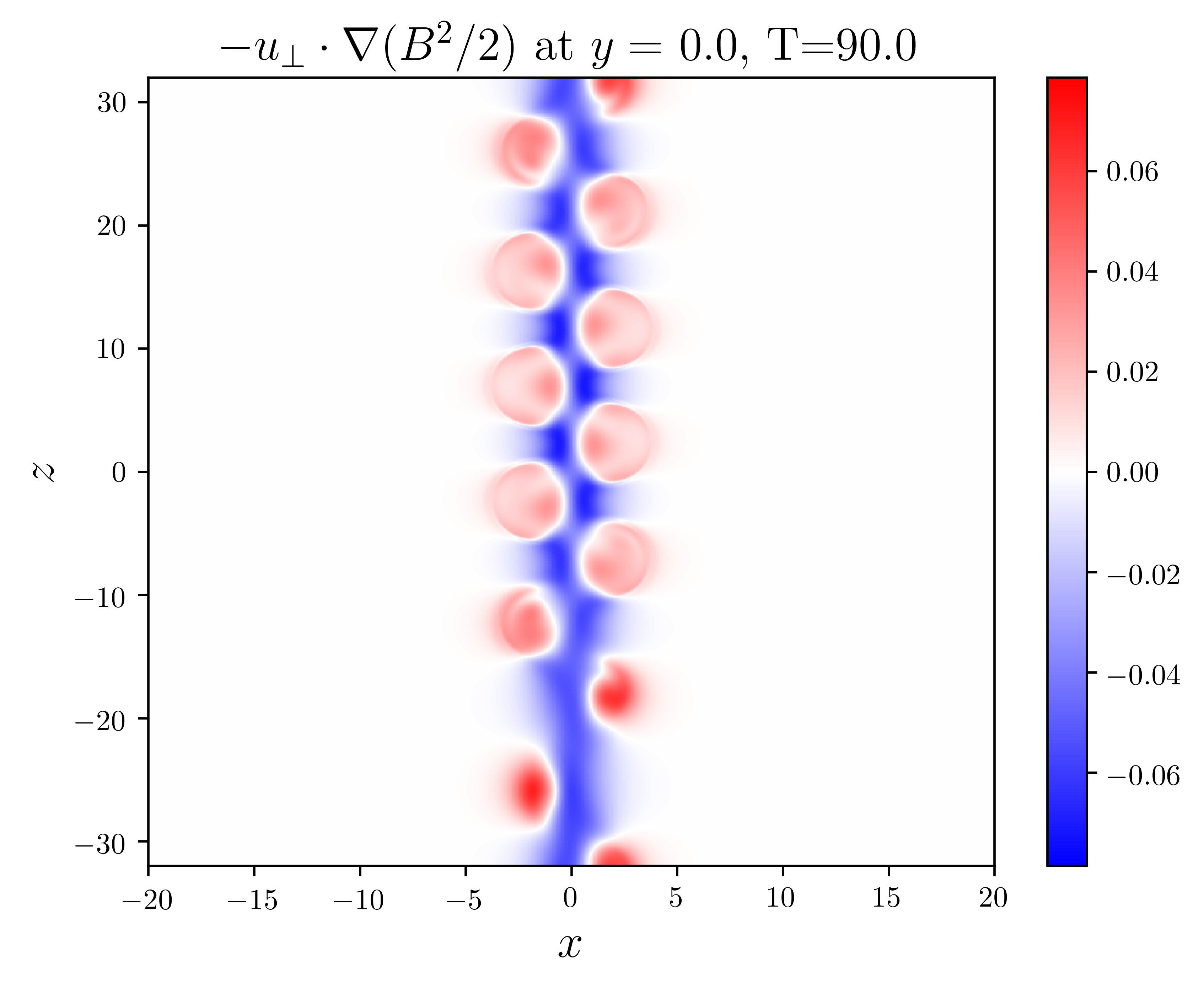}%
  \includegraphics[width=0.33\linewidth]{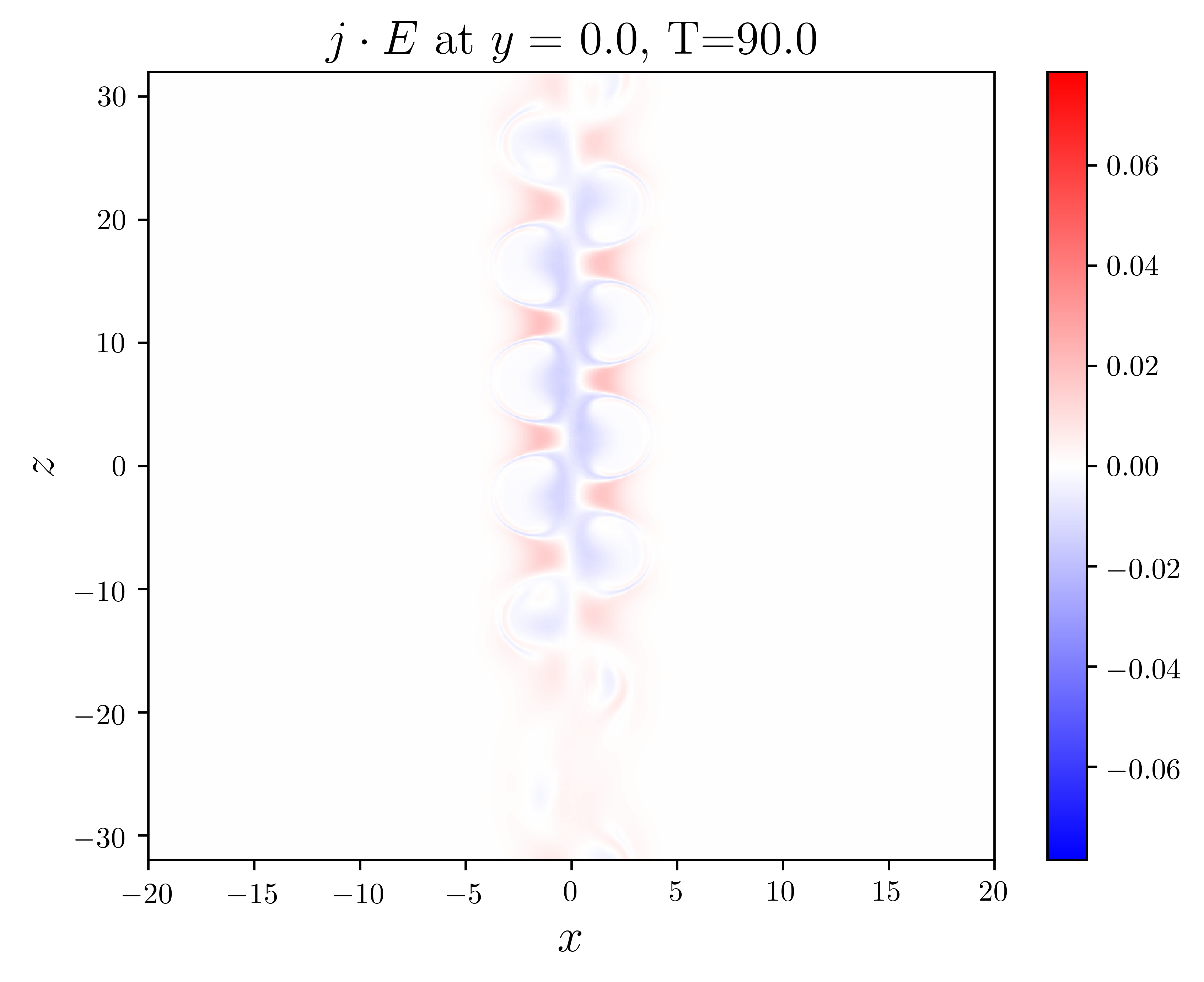}
  \includegraphics[width=0.33\linewidth]{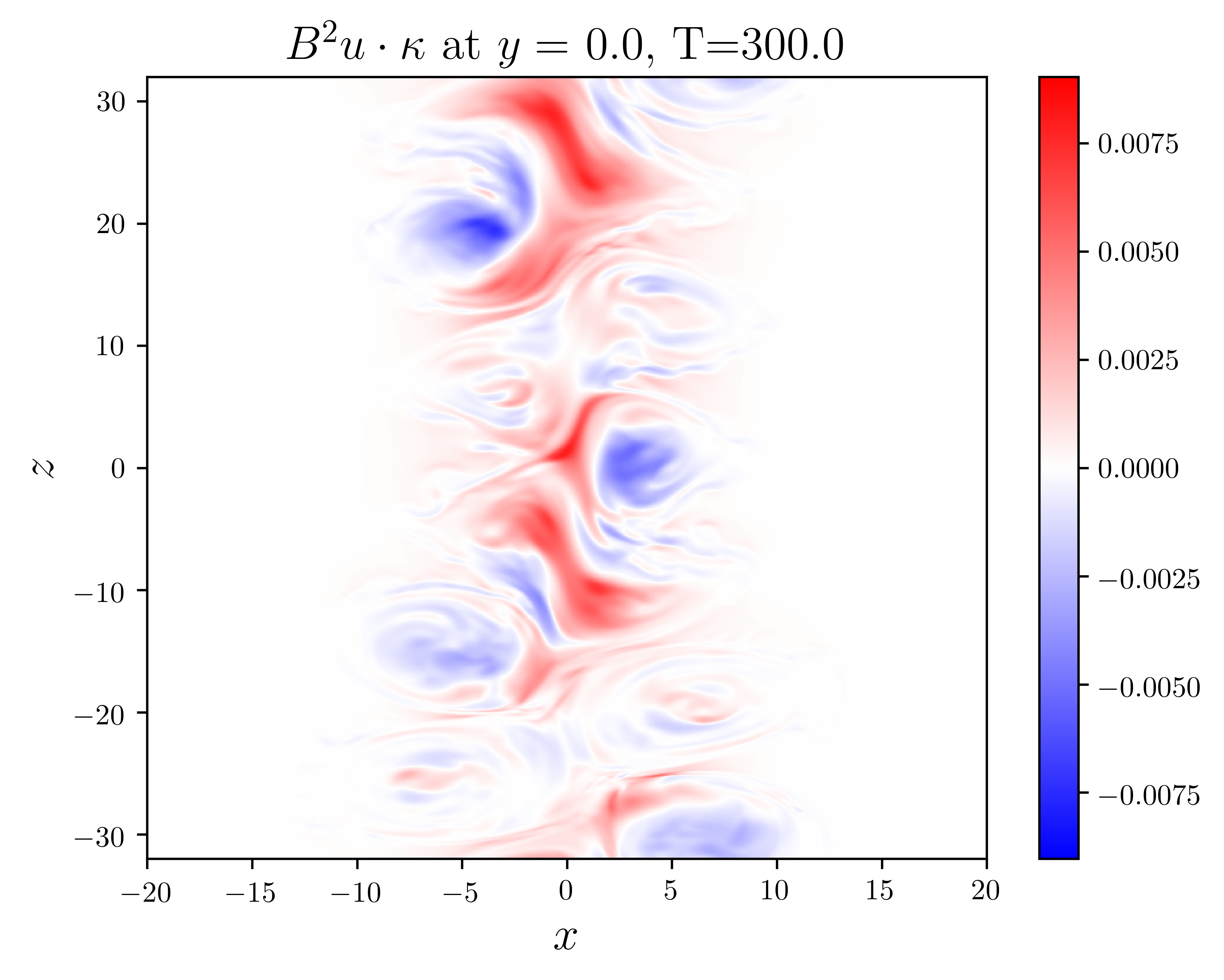}%
  \includegraphics[width=0.33\linewidth]{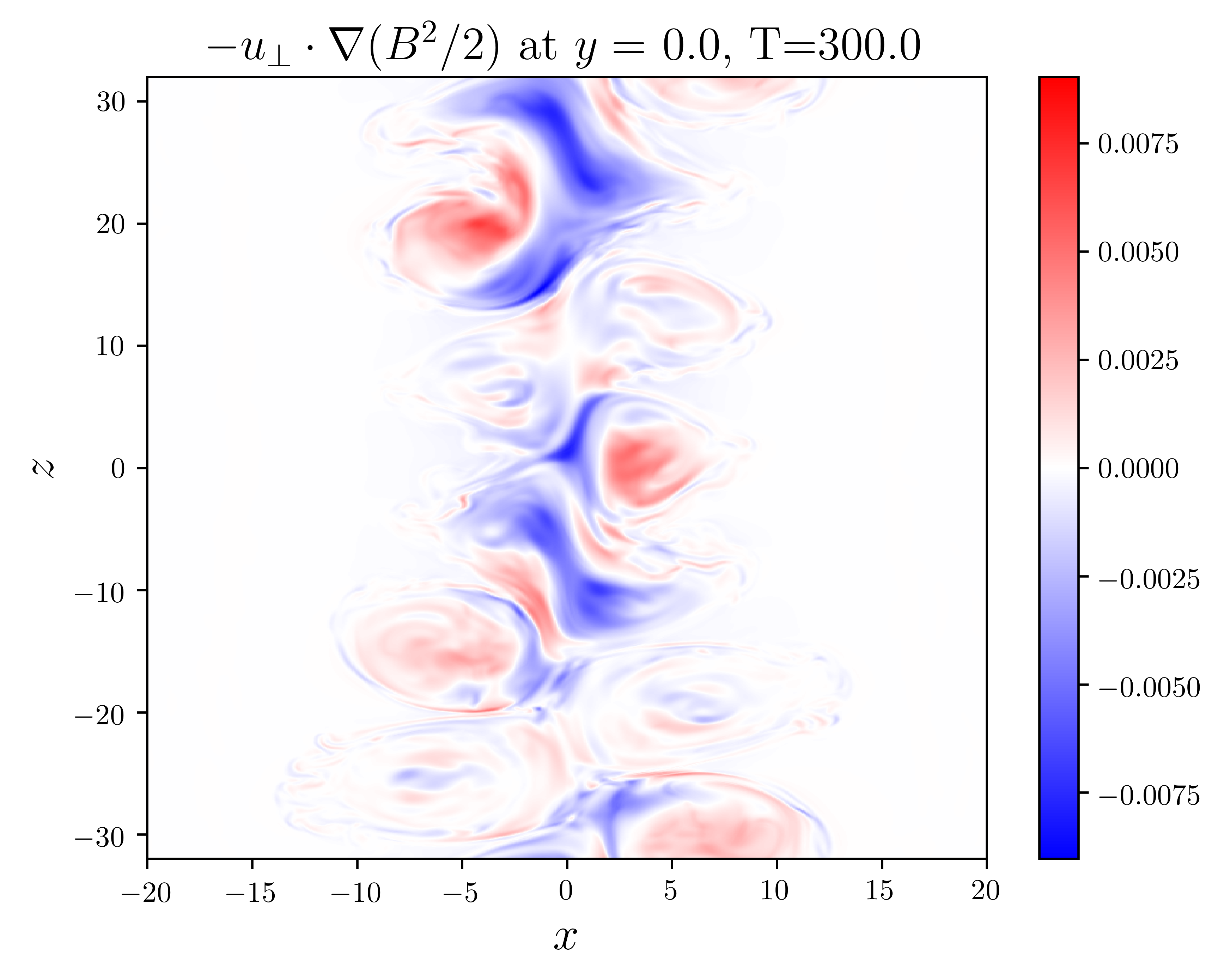}%
  \includegraphics[width=0.33\linewidth]{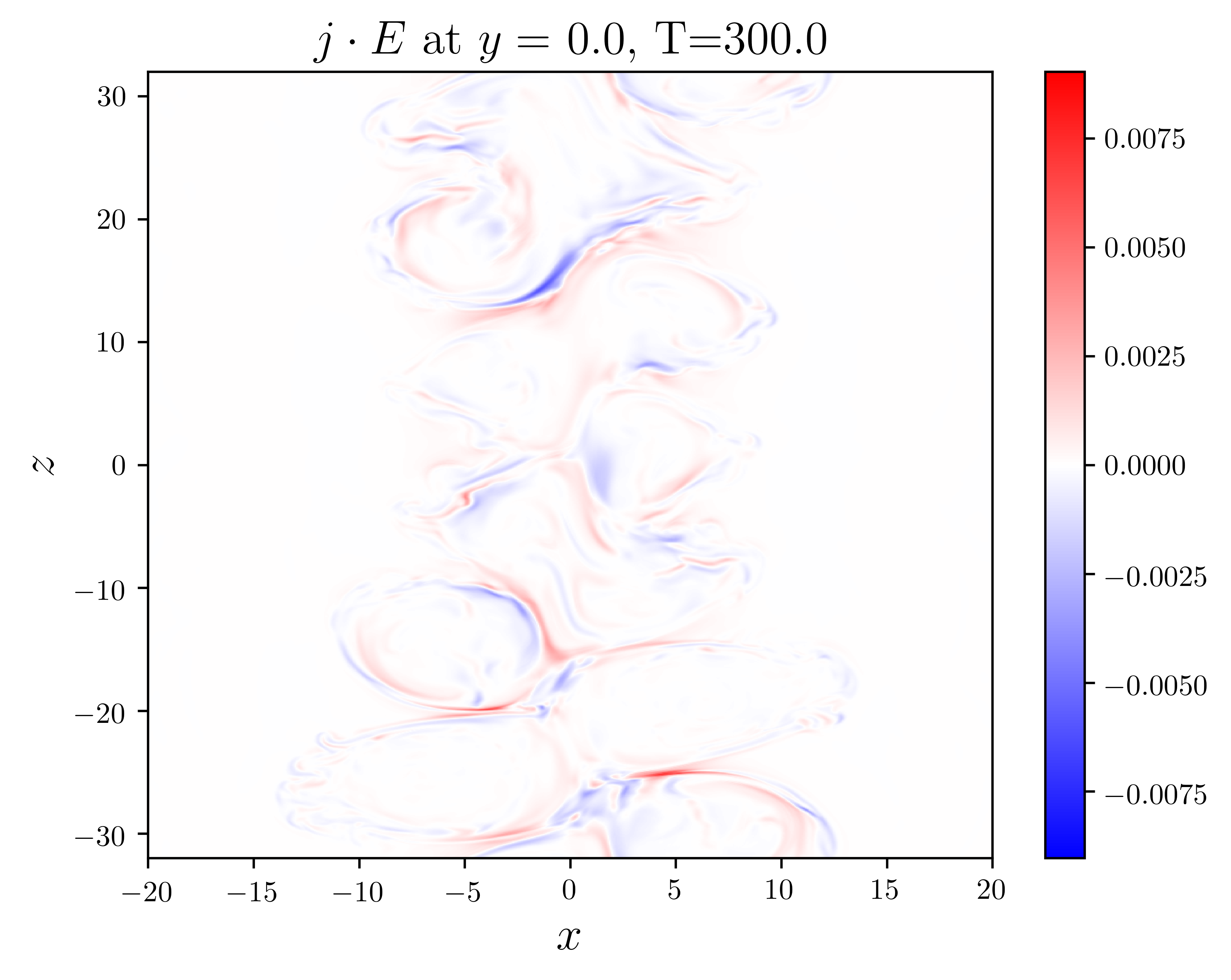}
  \caption{Similar to Figure \ref{fig:spatial1}, spatial distribution of the CR (left) and PE (middle) terms and their sum (right) in the $x$-$z$ plane at early ($t = 90$) and late ($t = 300$) times of the kink simulation. The spatial variation along the jet direction
comes from the superposition of modes with different wavelengths. }\label{fig:spatial2}
\end{figure}

Similar to Figure \ref{fig:spatial1}, we plot the spatial distribution of 
the conversion terms for the kink simulation, shown in Figure \ref{fig:spatial2}.
The spatial anti-correlation of the curvature and gradient terms is strong during
the initial unstable stage where there is very little net conversion of magnetic energy
(the total  $\boldj\cdot\boldE$ is nearly zero).
While more turbulent patterns develop at the later stage, the anticorrelation
between the two term is still clear.
Although the simulation is not run long enough, we expect that
the system will become more turbulent as it evolves further in time.

\section{Summary and conclusions}\label{sec:summary}

In this paper, the conversion of magnetic energy is 
discussed in the MHD framework. Both theoretical analysis and numerical 
simulations are presented, illustrating different mechanisms for the energy transfer. 
To summarize, our work yields the following conclusions: 

\begin{enumerate}
\item In the ideal MHD description, the total energy transfer from magnetic energy to 
plasma energy $\boldj\cdot\boldE$ can be separated into two dominant processes. 
Physically, the first part represents the relaxation of magnetic field line tension 
when the plasma flow velocity is aligned with the magnetic field curvature (the CR term); 
and the second part represents the situation where the perpendicular 
magnetic pressure gradients are anti-aligned with the plasma flow (the PE term).
\item 
We present several ideal MHD simulations for two types of systems: 
magnetic reconnection and kinked jet.
We find that, for both 3D reconnection and kinked jet, the CR and PE terms make
comparable contributions to the conversion of magnetic energy to plasma energy,
with the PE process playing a more important role overall in the magnetic energy conversion.
\item For both 3D reconnection and kinked jet, which have an inherent instability 
(tearing and kink)
to start the magnetic energy conversion process, a quasi-turbulent state is developed
as the system evolves further into the nonlinear stage. The relative importance of the CR and PE terms
appear to reverse in the turbulent stage compared to the linear stage.
For 3D reconnection, the CR term dominates the conversion of magnetic energy to plasma energy
in the beginning while the PE term becomes more important in the later more turbulent stage.
For kinked jet, the PE term dominates the magnetic energy conversion at the beginning
but the CR term becomes more important later.
\item In the 3D reconnection situation, the relative importance of the CR and PE terms
depend on the presence of guide field, most likely due to the impact of the guide field
on the strength of perpendicular expansion owing the magnetic pressure gradients. 
For a finite guide field ($B_g = 0.2$), we find that the CR term is overall the more important term. 
\end{enumerate}

These findings are potentially important for understanding particle energization
in 3D and over many nonlinear times.  
The role and importance of the PE term might have been under appreciated previously,
particularly given its overall dominance in producing substantially 
more magnetic energy conversion than the CR term. It will be interesting
to explore more in depth its consequences
on particle energization and transport processes in 3D magnetically dominated
systems.

\begin{acknowledgements}
We acknowledge the support from a NASA/LWS 
project 80HQTR20T0027. 
S.D. and H.L. also acknowledge the support by DOE OFES program 
and LANL/LDRD program.
Useful discussions with X. Li and F. Guo are gratefully acknowledged. 
Simulations were carried out using LANL's Institutional Computing resources.
\end{acknowledgements}

\bibliographystyle{aasjournal}
\bibliography{compression}

\appendix

\section{Relating Magnetic Energy Conversion to Plasma Energization}

The relationship between the CR+PE processes as described in 
Equation \eqref{eq:jdote2} and the plasma energization processes
remains, however, relatively unexplored. 
It is tempting to seek any correspondence between the magnetic 
energy conversion terms expressed in Equations \eqref{eq:jdote2} and
\eqref{eq:jdote3} with particle energization processes, particularly
in the context of the recent kinetic studies \citep[e.g.,][]{Li2017, Li2018}.
We now discuss some possible connections. 

\subsection{Single particle drifts and energization}

\citet{Beresnyak2016} have suggested that the particle's curvature drift 
acceleration could be related to the CR term discussed in 
Equation \eqref{eq:jdote2}. 
The curvature drift velocity of a single particle is
\begin{equation}
  \boldv_c = c\frac{2\mathcal{E}_\parallel}{qB}(\boldb\times\boldkappa),
\end{equation}
where $\mathcal{E}_\parallel = mv_\parallel^2 / 2$ is the 
particle parallel energy. 
Using the convective electric field, 
particle acceleration by curvature drift can be obtained 
by calculating the work done by the electric field on the 
particle's curvature drift velocity,
\begin{equation}\label{eq:curv_acc}
  \left.\frac{d\mathcal{E}}{dt}\right|_c = q\boldE\cdot\boldv_c = 2\mathcal{E}_\parallel \boldu\cdot\boldkappa.
\end{equation}
This process was emphasized in \citet{Beresnyak2016}. 
Similarly, using the grad-B drift velocity,
\begin{equation}
  \boldv_g = c\frac{\mathcal{E}_\perp}{qB}\frac{\boldB\times\nabla B}{B^2},
\end{equation}
we find that particle acceleration by grad-B drift as,
\begin{equation}\label{eq:grad_acc}
  \left.\frac{d\mathcal{E}}{dt}\right|_g = q\boldE\cdot\boldv_g = \frac{\mathcal{E}_\perp}{B^2}\boldu\cdot\nabla_{\perp}\kl{\frac{B^2}{2}}.
\end{equation}
This grad-B drift acceleration term was not included in 
\citet{Beresnyak2016}, as they primarily focused on the
incompressible MHD limit where the curvature drift term dominates. 
In a general compressible system, the conclusions 
of \citet{Beresnyak2016} need to be revisited and expanded.

Equations \eqref{eq:curv_acc} and \eqref{eq:grad_acc} contain 
terms $\boldu\cdot\boldkappa$ and $\boldu\cdot\nabla_{\perp}\kl{{B^2}/{2}}$,
respectively, which are included in the two magnetic energy conversion
processes of CR and PE as shown in Equation \eqref{eq:jdote2}. 
We have placed these expressions side-by-side for comparison,
\begin{eqnarray}
 \boldj\cdot\boldE &=& \frac{B^2}{4\pi}\boldu\cdot\boldkappa - \boldu\cdot\nabla_{\perp}\kl{\frac{B^2}{8\pi}}, ~~{\rm magnetic ~energy~ conversion~rate~in~MHD~ (complete)}\nonumber\\
   \frac{d\mathcal{E}}{dt} &\ \sim& 2\mathcal{E}_\parallel \boldu\cdot\boldkappa +\frac{4\pi \mathcal{E}_\perp}{B^2}\boldu\cdot\nabla_{\perp}\kl{\frac{B^2}{8\pi}},
   ~~{\rm single~particle ~energy~ change~rate~(partial)}
 \label{eq:jdote1}
\end{eqnarray}

One has to be very cautious in drawing any conclusions from this seeming
similarity. For example, 
from a single particle perspective, the particle energy gain associated 
with curvature drift ($d\mathcal{E} / dt |_c > 0$) has the same sign as 
the CR process with
an increase in plasma energy ($\boldj\cdot\boldE > 0$). 
However, the particle energy gain associated with gradient drift 
($d\mathcal{E}/dt|_g > 0$) has the {\it opposite} sign as 
the PE process with a {\it decrease} in plasma 
energy ($\boldj\cdot\boldE < 0$). So, we should not 
directly relate the CR and PE processes in magnetic energy 
conversion with the curvature
drift and grad-B drift processes in particle energization. 
This is not surprising --- an example is that particles gain energy 
due to gradient drift at a perpendicular shock while the 
magnetic field energy increases at the shock front.
Both particles and magnetic fields gain energy from the bulk flow energy.
For the acceleration of energetic particles, 
Equations \eqref{eq:curv_acc} and \eqref{eq:grad_acc} 
show that the energy gained by particles is proportional to the 
particle (parallel or perpendicular) energy. Therefore, 
high-energy particles may gain a significant amount of energy 
via a Fermi-like mechanism upon encountering a region 
with $\boldu\cdot\boldkappa > 0$ or $\boldu\cdot\nabla_{\perp}B^2 > 0$.
This is different from the conversion of magnetic energy in 
Equation \eqref{eq:jdote2}, which does not depend on particle energy.

\subsection{Particle drift currents and energization}

Using the kinetic studies of magnetic reconnection, 
\citet{Li2018} showed from the Vlasov equation that, in the
limit when particles are sufficiently magnetized, 
the particle energization due to perpendicular electric field can be expressed 
as the sum of curvature drift, gradient drift, perpendicular 
magnetization, and bulk acceleration (the inertial term) 
in the limit of gyrotropic pressure. The curvature and gradient 
drift terms are found to be dominant over the others \citep{Li2017, Li2018}.
Specifically, 
the energization of particles due to drifts can also be
calculated as
\begin{eqnarray}
  \boldj_c \cdot \boldE &=& 
  p_{\parallel}c\frac{\boldb\times\boldkappa}{B} \cdot \boldE = p_{\parallel} \boldu \cdot \boldkappa; \label{eq:jcde}\\
  \boldj_g \cdot \boldE &=& 
  ( p_{\perp}c\frac{\boldB}{B^3}\times\nabla B) \cdot \boldE = \frac{p_{\perp}}{B^2} \boldu\cdot\nabla_{\perp}\left(\frac{B^2}{2}\right), \label{eq:jgde}
\end{eqnarray}
where $p_{\parallel}$ and $p_{\perp}$ represent parallel and perpendicular pressure.
Fully kinetic PIC simulations have shown that 
curvature drift acceleration is found to be the dominant 
particle acceleration mechanism, while grad-B drift typically 
has the effect of decelerating particles.

\begin{figure}[!htp]
  \centering
  \includegraphics[width=0.5\linewidth]{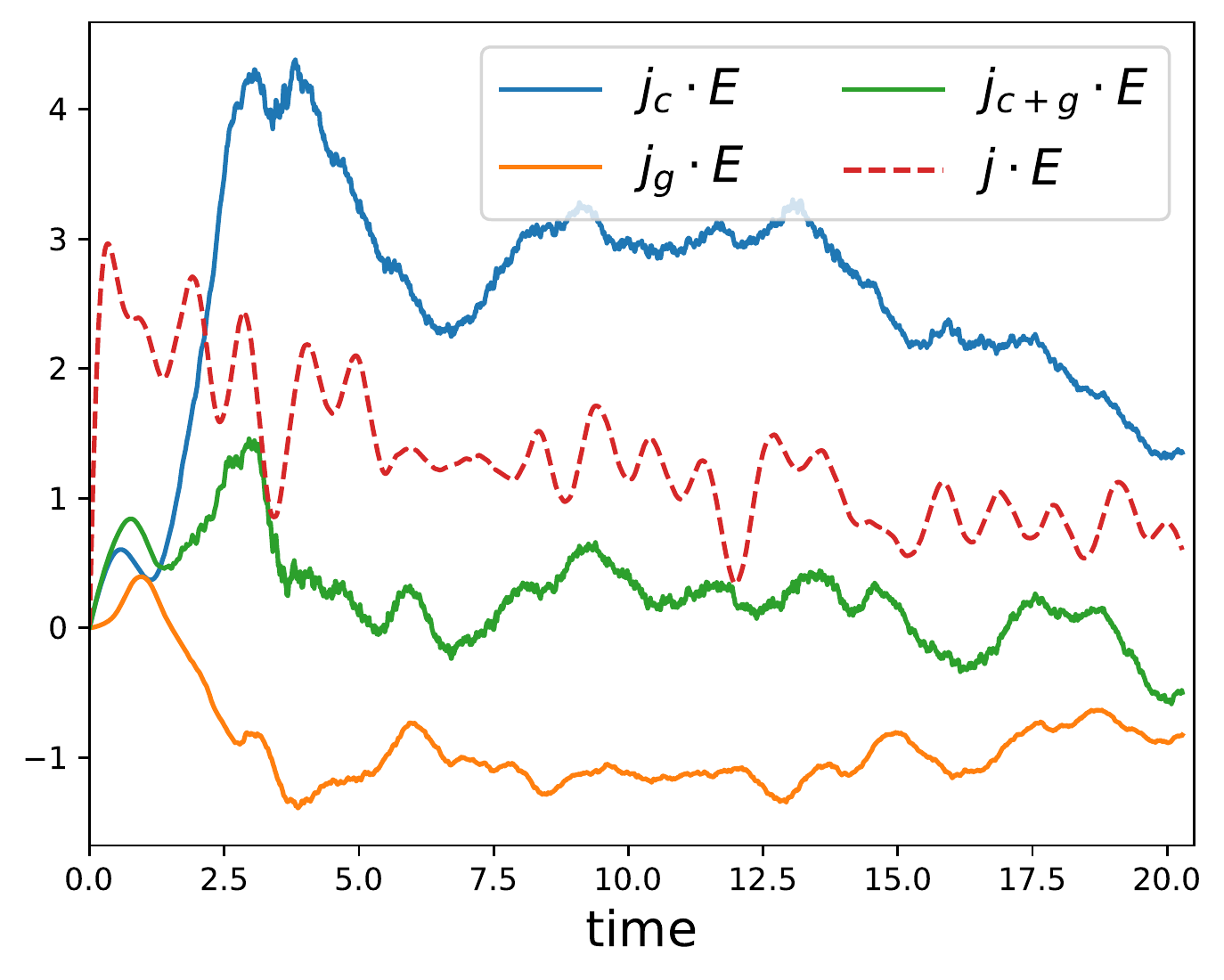}
  \caption{Time history of particle-based energization terms in Equation \eqref{eq:jcde} and \eqref{eq:jgde} for the magnetic reconnection simulation. The sum of the two terms is plotted as the green curve ($\boldj_{c+g}\cdot \boldE$). The total magnetic energy conversion $\boldj\cdot\boldE$ is plotted as the red dashed line for comparison.}\label{fig:energy1-part}
\end{figure}

By assuming an isotropic pressure, we could use our 3D MHD reconnection
simulation described in \S \ref{sec:reconnection} to study
the evolution of Equations \eqref{eq:jcde} and \eqref{eq:jgde}. 
Figure \ref{fig:energy1-part} shows the plasma energy change rates
due to curvature and gradient drift currents 
using Equation \eqref{eq:jcde} and \eqref{eq:jgde} 
assuming an isotropic pressure.
The figure shows clearly that the plasma energy gain via
curvature drift  $\boldj_c\cdot\boldE$ remains positive 
throughout the simulation, while the plasma energy change via the 
grad-B drift process $\boldj_g\cdot\boldE$ is negative for most of the time.
The results here are qualitatively similar to those from 
PIC simulations \citep{Li2017}, indicating that the particle 
acceleration is perhaps more sensitive to the global field 
structure rather than the detailed kinetic physics near 
the reconnection X-line.
We have also plotted the total $\boldj \cdot \boldE$ (red dashed line)
as given in Equation \eqref{eq:jdote2}. Note that it exhibits
difference from the sum of Equations \eqref{eq:jcde} and \eqref{eq:jgde} (green curve),
indicating that these two processes do not capture the total
magnetic energy conversion.

It should be emphasized that the two terms shown in 
Equations \eqref{eq:jcde} and \eqref{eq:jgde} should not be 
equated to the CR and PE terms in Equation  \eqref{eq:jdote2}. 
For example, the particle gain via 
the curvature drift current is proportional to the particle 
(parallel) pressure whereas the CR magnetic energy conversion 
term is completely independent of the particle pressure.
Intuitively, however, the two descriptions appear to be linked.
For a plasma that is approximately in 
pressure balance, the particle pressure is anti-correlated 
with the magnetic pressure, so that particles tend to gain 
a large amount of energy in the high-curvature region 
where the particle pressure tends also to be high.

As for the grad-B drift term and the PE process, 
one needs to keep in mind that there is no direct correlation or
correspondence between the particle energy change via grad-B drifts and
the magnetic energy change. For example, a decrease in 
the bulk kinetic flow energy could cause a positive 
$\boldu\cdot\nabla_{\perp}\left({B^2}/{2}\right)$
so that both the particle energy gain via grad-B and the magnetic energy 
increase can occur.

\end{document}